\documentclass[twocolumn,preprintnumbers,superscriptaddress,amsmath,amssymb,10pt,a4paper,nofootinbib]{revtex4-1}

\usepackage{geometry}
\usepackage{mathtools}
\usepackage{graphicx}
\usepackage{dcolumn}
\usepackage{bm}
\usepackage{makecell}
\usepackage{xcolor}
\usepackage{enumerate}
\usepackage[caption=false]{subfig}
\usepackage{float}
\usepackage[unicode=true,pdfusetitle,
 bookmarks=true,bookmarksnumbered=false,bookmarksopen=false,
 breaklinks=false,pdfborder={0 0 1},backref=false,colorlinks=true]
 {hyperref}
\hypersetup{linkcolor=blue,urlcolor=blue,citecolor=blue}
\usepackage{cleveref}
\usepackage{physics}

\newcommand{\be}{\begin{equation}}
\newcommand{\ee}{\end{equation}}
\newcommand{\bd}{\begin{displaymath}}
\newcommand{\ed}{\end{displaymath}}
\newcommand{\BE}{\begin{eqnarray}}
\newcommand{\EE}{\end{eqnarray}}

\newcommand{\avg}[1]{\left\langle{#1}\right\rangle}

\begin{document} 

\title{Niche overlap and Hopfield-like interactions in generalised random Lotka--Volterra systems}

\author{Enrique Rozas Garcia}
\email{enrique.rozas.garcia@physics.gu.se}
\affiliation{Department of Physics, Gothenburg University, 41296 Gothenburg, Sweden}
\affiliation{Instituto de F\'isica Interdisciplinar y Sistemas Complejos, IFISC (CSIC-UIB), Campus Universitat Illes Balears, E-07122 Palma de Mallorca, Spain}
\author{Mark J. Crumpton}
\email{mark.j.crumpton@kcl.ac.uk}
\affiliation{Department of Mathematics, King's College London, London WC2R 2LS, United Kingdom}
\affiliation{Department of Physics and Astronomy, School of Natural Sciences, The University of Manchester, Manchester M13 9PL, UK}

\author{Tobias Galla}
\email{tobias.galla@ifisc.uib-csic.es}
\affiliation{Department of Physics and Astronomy, School of Natural Sciences, The University of Manchester, Manchester M13 9PL, UK}
\affiliation{Instituto de F\'isica Interdisciplinar y Sistemas Complejos, IFISC (CSIC-UIB), Campus Universitat Illes Balears, E-07122 Palma de Mallorca, Spain}
\date{\today}

\begin{abstract}
We study communities emerging from generalised random Lotka--Volterra dynamics with a large number of species with interactions determined by the degree of niche overlap. Each species is endowed with a number of traits, and competition between pairs of species increases with their similarity in trait space. This leads to a model with random Hopfield-like interactions. We use tools from the theory of disordered systems, notably dynamic mean field theory, to characterise the statistics of the resulting communities at stable fixed points and determine analytically when stability breaks down. Two distinct types of transition are identified in this way, both marked by diverging abundances, but differing in the behaviour of the integrated response function. At fixed points only a fraction of the initial pool of species survives. We numerically study the eigenvalue spectra of the interaction matrix between extant species. We find evidence that the two types of dynamical transition are, respectively, associated with the bulk spectrum or an outlier eigenvalue crossing into the right half of the complex plane.
\end{abstract}

\maketitle

\section{Introduction}
The foundations of the theory of disordered systems date back close to 50 years \cite{edwards1975theory}. Initially, the aim was to understand certain magnetic states in condensed matter physics (`spin glasses') \cite{mezard1987spin}. However, it became clear that applications of the tools developed for disordered systems had a reach far beyond the boundaries of physics. Methods such as replica theory or dynamic generating functionals were quickly adapted and used to answer questions in neural networks \cite{coolen_stat, Coolen_dyn,coolen_kuehn_sollich}, to study the Minority Game \cite{coolen2005mathematical} (sometimes presented as a simple model of a financial market), or indeed evolutionary bi-matrix games and so-called Nash equilibria \cite{berg}.

The defining feature of disordered systems is the presence of quenched disorder. That is, the system is made up of many constituents, and the interactions between these are determined by coefficients that are drawn at random at the beginning, but then remain fixed as the dynamics of the system unfolds. The disorder leads to complicated energy landscapes. The number of local minima can grow exponentially in the size of the system, and is often organised in a hierarchical manner. Dynamic phenomena in disordered systems include ergodicity breaking and so-called ageing \cite{mezard1987spin,hertz}.

Ideas and methods from the physics of disordered systems have also been used to study complex ecosystems \cite{may1972will, allesina2012stability, opper1992phase, yoshino1, yoshino2, bunin2016interaction, galla2018dynamically, biroli2018marginally, altieri2021properties}. The word `complex' in this context indicates that the ecosystem is composed of a large number of species, and that these species are subject to randomly drawn interaction coefficients. In this paper we continue this line of work, and focus on a Lotka--Volterra system with Hopfield-like interactions \cite{coolen_stat,Coolen_dyn,coolen_kuehn_sollich}. More specifically, we are interested in a set of $N$ species ($N\gg 1$), whose abundances develop in time following a generalised Lotka--Volterra equation (details will follow in Sec.~\ref{sec:model_definitions}). This involves an $N\times N$ matrix $a_{ij}$ of interaction coefficients. Existing work on the statistical physics of complex ecosystems has mostly focused on the case in which the interaction matrix is drawn from distributions with either no correlations between different matrix elements, or only correlations between diagonally opposed entries \cite{may1972will,opper1992phase,bunin2016interaction,biroli2018marginally,galla2018dynamically,roy2019numerical, altieri2021properties, Altieri2022}. There is also work on cases in which the matrix is composed of blocks, and where the elements in different blocks have different statistics \cite{Poley}. One common element shared by many existing random Lotka--Volterra models is that the finest level of modelling is set by the interaction coefficients. No further assumptions are made about the properties of the species, and how the species interactions come about from these properties.

The Hopfield model is inspired by structures first used in neural networks \cite{hopfield1982neural, hebb-1949, coolen_kuehn_sollich}. Translated into the language of ecology, the starting point is now a set of species and a set of traits. Each species can either possess or not possess a given trait. This assignment of traits to species, in turn, determines how species will interact. Broadly speaking, the interaction between two species will be more competitive the more traits they share (i.e. the more similar the two species are). This type of interaction structure has also been studied in models combining resources and consumers, both in economics and in ecology \cite{macarthur1955fluctuations, demartino2006, yoshino1, yoshino2, advani2018}. A particularly notable model is that by MacArthur and collaborators \cite{MacArthur1967,MacArthur1970,Cheeson1990}. Analyses of random replicator systems with `Hebbian' interactions \cite{galla_hebbian} have shown interesting statistical mechanics, and in particular types of phase transition that are different from those seen in replicator systems with Gaussian couplings.

In this paper, we set out to characterise the behaviour of a Lotka--Volterra system with Hopfield-like interactions, where we allow for a degree of `mild' dilution (the system is not fully connected, but each species still interacts with an extensive number of other species). A system of replicator equations with such interactions was studied in \cite{galla_hebbian}. Our aim is to calculate the statistics of fixed points in the phase where such fixed points are attained and identify the onset of instability. As in the system with Gaussian interactions, we find that only a proportion of the initial species survive at stable fixed points. Recent work \cite{baron2022non} on Gaussian systems has shown that the reduced interaction matrix (the matrix of interaction coefficients among the surviving species) has intricate statistics. Specifically, its bulk and outlier eigenvalues can be related to different types of dynamic phase transitions. As we will show, the types of phase transition seen in our model differ from those in the Gaussian model. One aim of the current paper is therefore to establish (in simulations) how these transitions relate to the spectra of the interaction matrix of the extant species.

The remainder of the paper is organised as follows. In Sec.~\ref{sec:model_definitions} we define the model and introduce the necessary notation. Sec.~\ref{sec:gf_analysis} then contains the mathematical analysis. This is based on so-called `generating functionals' and dynamic mean field theory. The phase diagram and further behaviour of the model are then discussed in Sec.~\ref{sec:pg}. In Sec.~\ref{sec:reduced_matrix} we finally turn to a study of the spectra of the reduced interaction matrix and their relation to the phase diagram. We conclude the paper with a discussion and an outlook in Sec.~\ref{sec:concl}.

\section{Model definitions}
\label{sec:model_definitions}
	
We will study the following generalised Lotka--Volterra equation (gLVE)
\begin{equation}
\dot{x}_i(t) = x_i(t)\left[K_i-u_ix_i+\sum_{j\neq i}c_{ij}J_{ij}x_j\right], \label{HLV}
\end{equation}
where the $x_i\geq0$ represent the abundances (or population densities) of different species, $i=1,\dots N$. We always assume  initial conditions for which all $x_i$ are strictly positive. 

The quantities $u_i>0$ denote the strength of intraspecific competition, and the $a_{ij}=c_{ij}J_{ij}$ represent the interspecific interactions. The $K_i$ (together with the $u_i$) set the carrying capacities of the species in the absence of interactions between different species ($x_i$ then tends to $K_i/u$ in the long run). We focus on the case $u_i\equiv u$ for all $i$, noting that $u$ controls the time scale on which the non-interacting system approaches the fixed point $x_i\equiv K_i/u$. We allow for general positive values of $u$ throughout our analysis, but in an effort to keep the number of parameters manageable we set $K_i\equiv 1$.

The dilution variables $c_{ij}\in\{0,1\}$ ($i\neq j$) determine which species interact with one another, i.e. they set the topology of the interaction network. For each pair $i<j$, the coefficients $c_{ij}$ and $c_{ji}$ are chosen from a Bernoulli distribution with
\begin{equation}
    \langle c_{ij}\rangle = c, \qquad \langle c_{ij}c_{ji}\rangle - c^2 = \Gamma c(1-c).   \label{prop_c}
\end{equation}
We thus have $P(c_{ij}=1)=c$ for all $i\neq j$, i.e. $c$ is the analog of what May called `connectance' \cite{may1972will}. The parameter $\Gamma$ is restricted to the range from $-1$ to $1$ by construction, but we note that not all choices of pairs $(c,\Gamma)$ are possible (see Supplemental Material \cite{Supplement} for details). We note that the choice of the diagonal coefficients $c_{ii}$ is irrelevant as we set $J_{ii}=0$ below

Throughout our paper, $c$ is chosen not to scale with $N$ [$c={\cal O}(N^0)$]. This means that each species interacts with an ${\cal O}(N)$ number of other species. We are therefore not studying a `dilute' system in the sense of random matrix theory. The extensive connectivity allows us to use established methods from dynamic mean-field theory. Truly dilute systems with $c={\cal O}(1/N)$ (and where consequently each species only interacts with a finite number of other species) can be expected to behave very differently, see e.g. \cite{Marcus2022}, and a theoretical analysis would be much more intricate.

Interaction links in our system are directed, that is, an effect of the presence of species $j$ on the dynamics of $i$ does not necessarily imply the reverse. The parameter $\Gamma$ measures the correlations between $c_{ij}$ and $c_{ji}$. A choice of $\Gamma=-1$ implies $c_{ij}=1-c_{ji}$ with probability one, and $\Gamma=1$ means that $c_{ij}=c_{ji}$ with probability one.

The matrix $J_{ij}$ determines the strength of the effects of the presence of species $j$ on the dynamics of the abundance of species $i$. Positive values of $J_{ij}$ imply that the population of species $j$ is beneficial to the growth of species $i$, while negative values imply a detrimental interaction. In ecological terms the signs of the pair of interactions $(\text{sgn}\,J_{ij}, \text{sgn}\,J_{ji})$ determine whether two species are in a mutualistic relation $(+,+)$, whether they compete with one another $(-,-)$, or whether there is an antagonistic predator-prey relation between $i$ and $j$ $(\pm,\mp)$ \cite{allesina2012stability}.

In this work, the interspecific interaction is chosen according to a niche overlap heuristic (see e.g. \cite{KNEITEL20081731}). We assume each species is described by a set of binary traits, labelled $\mu=1,\dots, P$, and that a pair of species will compete in proportion to the similarity between the two species (i.e, the number of traits which both or neither species possess). We write  $\xi^\mu_i=+1$, if species $i$ has trait $\mu$, and $\xi_i^\mu=-1$ if the species does not possess the trait. Interactions are then assumed to be of the form
\begin{equation}
J_{ij} = \begin{dcases}
-\frac{1}{cN}\sum_{\mu=1}^{\alpha c N}\xi_i^\mu\xi_j^\mu & i\neq j \\
0 & i=j
\end{dcases} \ .
\label{coupling} 
\end{equation}
We have here set $P=\alpha c N$, with $\alpha>0$ a model parameter (in simulations $P$ is restricted to integer values.) That is to say, we assume that the number of traits is proportional to the number of species in the system. The interaction in Eq.~\eqref{coupling} is reminiscent of the Hopfield model, used in the context of neural networks \cite{Coolen_dyn, hopfield1982neural}. This suggests interesting phase behaviour when $\alpha={\cal O}(N^0)$, which is the regime we focus on. We have normalised the interaction strength by $cN$, the mean number of species that any one species will interact with. We will refer to the random variables $c_{ij}$ and $\xi^\mu_i$ as the \textit{disorder} of the system. 

The traits $\xi_i^\mu$ are chosen to be $\xi_i^\mu=\pm 1$ with equal probability, and there is no correlation between the different $\xi_i^\mu$. This implies that the distribution of the $J_{ij}$ approaches a Gaussian as $N\to\infty$, reminiscent of the model studied for example in \cite{bunin2016interaction, galla2018dynamically}. We note, however, that the Hopfield structure introduces correlations between the different $J_{ij}$, which are different from the correlations studied in the earlier literature. We highlight again the structural similarity to MacArthur's consumer-resource model \cite{MacArthur1967, MacArthur1970, Cheeson1990} (see also  \cite{advani2018, demartino2006}  for statistical physics studies), noting though that the latter model is more sophisticated, with dynamical equations both for consumers and resources.

\section{Generating functional analysis and stability}\label{sec:gf_analysis}
  
\subsection{Generating functional and effective process}

We analyse the system in Eq.~(\ref{HLV}) using dynamic generating functionals, an established method in the theory of disordered systems \cite{de1978dynamics, Coolen_dyn}. This leads to an effective `dynamic mean field theory'. Similar approaches have been used to study Lotka--Volterra systems with Gaussian random couplings \cite{opper1992phase, galla2018dynamically, sidhom2020ecological}. We note that an alternative approach is based on the so-called cavity method \cite{bunin2016interaction, bunin2017ecological, roy2019numerical}. We also add that the dynamics admit a Liapunov function when $c=1$ (leading to a symmetric interaction matrix). Methods from the equilibrium statistical physics of disordered systems such as the replica approach can be used in this special case (for examples see \cite{biroli2018marginally, altieri2021properties}).

The outcome of the application of these techniques is an effective stochastic process for a `representative species'. The ensemble of realisations of stochastic processes is statistically equivalent to the set of single-species trajectories $x_i(t)$ of the disordered dynamical system in Eq.~\eqref{HLV}. The dynamic mean-field description becomes exact in the thermodynamic limit (${N\rightarrow\infty}$). Overall, in this limit, the infinite-dimensional deterministic dynamical system in Eq.~(\ref{HLV}) is traded for an effective single-species process which is non-local in time (it involves retarded self-interaction) and contains coloured noise.

The generating functional analysis begins from 
\begin{equation}\label{eq:HLV_1}
\dot{x}_i(t) = x_i(t)\left[1-ux_i+\sum_{j\neq i}c_{ij}J_{ij}x_j - h_i(t)\right],
\end{equation}
where we have introduced the perturbation fields $h_i(t)$ in order to calculate linear response functions. These fields are not actually part of the model and are set to zero at the end of the calculation, as well as in all simulations shown in the paper. For more details see the Supplementary Material (SM). The generating functional of this dynamical system is given by
\begin{equation}\label{eq:gf}
\mathcal{Z}[\psi_i(t), h_i(t)] = \left\langle\exp\left(i\sum_i\int dt\, x_i(t)\psi_i(t)\right)\right\rangle_{\mbox{\footnotesize paths}},
\end{equation}
where the average is over paths $[x_1(t),\dots,x_N(t)]$ of the dynamics in Eq.~(\ref{eq:HLV_1}). The $\psi_i(t)$ constitute a source field. The generating functional in Eq.~(\ref{eq:gf}) is the Fourier transform of the probability measure in the space of paths generated by Eq.~(\ref{eq:HLV_1}).

The final outcome of the generating-functional analysis is a set of equations for the dynamic macroscopic order parameters of the problem. For the Lotka--Volterra model these are
\BE
M(t) &=& \lim_{N\rightarrow\infty}\frac{1}{N} \sum_{i=1}^N\overline{\langle x_i(t)\rangle_0}, \nonumber \\
C(t,t')&=& \lim_{N\rightarrow\infty}\frac{1}{N} \sum_{i=1}^N\overline{\langle x_i(t)x_i(t')\rangle_0}, \nonumber \\
G(t,t') &=& \lim_{N\rightarrow\infty}\frac{1}{N} \sum_{i=1}^N\frac{\delta\overline{\langle x_i(t)\rangle_0}}{\delta h_i(t')}, \label{theparameters}
\EE
where $\delta$ denotes a functional derivative and $\langle\cdots\rangle_0$ stands for an average over random initial conditions. The overbar $\overline{\,\cdots\,}$ represents the average over the disorder, i.e. over the $c_{ij}$ and $\xi_i^\mu$. The order parameters can be obtained from the disorder-averaged generating functional as derivatives with respect to the fields $\psi_i(t)$ and/or $h_i(t)$, evaluated at $\psi_i(t)\equiv 0$ and $h_i(t)\equiv 0$.

\begin{figure*}[t]
    \centering
    \includegraphics[width=1\linewidth]{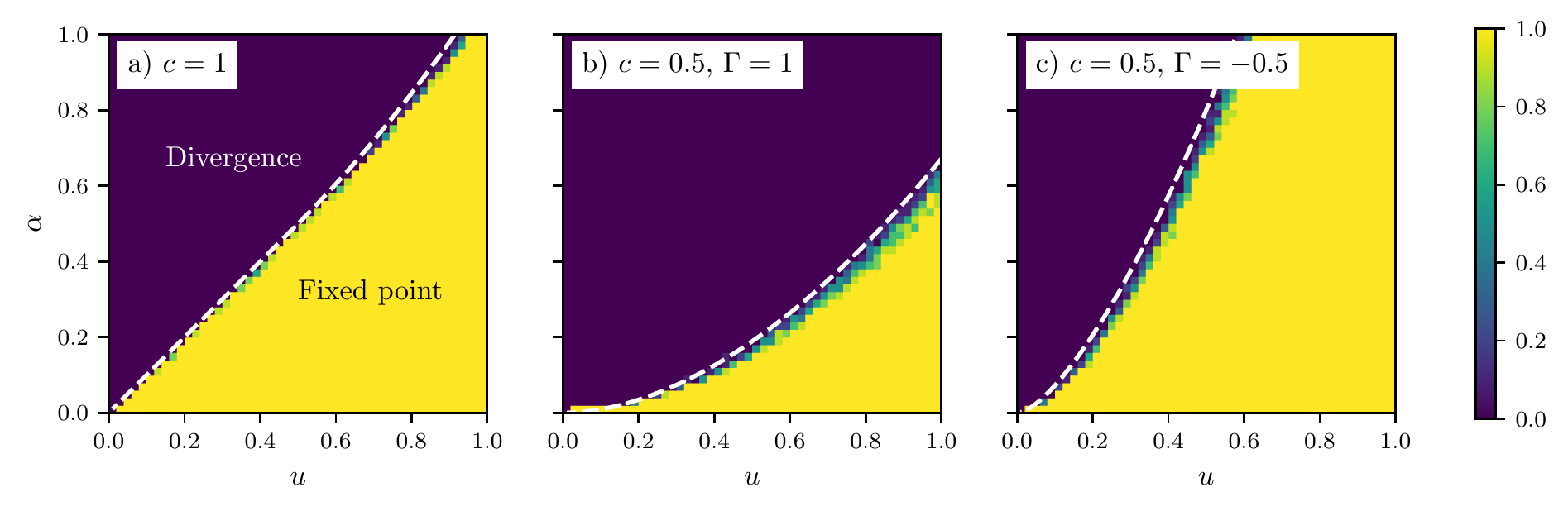}
	\caption{Fixed points and divergences of the dynamics in Eq.~\eqref{HLV}. Each panel illustrates the behaviour of the model for different choices of $c$ and $\Gamma$. The heatmap indicates the fraction of samples that converge to a fixed point after numerical integration of the gLVE. The criteria for the identification of convergence or divergence are described in Appendix \ref{app:numerics}. The dashed lines are theoretical predictions for the onset of divergence (see Sec.~\ref{sec:stab}).}
 	\label{fig:cmap}
\end{figure*}

The order parameters in Eqs.~(\ref{theparameters}) are determined self-consistently from an effective process for a single representative (`mean field') species. The procedure to derive the effective equations is well-documented  \cite{opper1992phase, coolen2000statistical, coolen2005mathematical, verbeiren2003dilution}, therefore, we only report the final result (a more detailed derivation can be found in the SM). The effective single-species process for the model is given by
\BE
\dot{x}(t)&=&x(t)\Biggr[1-ux(t)-\alpha\int_0^tdt'[cG(\mathbb{I}-G)^{-1}
\nonumber \\
&&+\Gamma(1-c)G](t,t')x(t')-\eta(t)\Biggr], \label{effective_dynamics}
\EE
where $\mathbb{I}$ is the identity operator and $\eta(t)$ is coloured Gaussian noise with zero mean and correlations in time, given by 
\BE
\langle \eta(t)\eta(t')\rangle &=& \alpha[c(\mathbb{I}-G)^{-1}C(\mathbb{I}-G^T)^{-1} \nonumber \\
&&+(1-c)C](t,t'). \label{eff_noise}
\EE

The order parameters in Eqs.~\eqref{theparameters} are to be obtained self-consistently from the following expressions,
\BE
M(t) &=& \langle x(t)\rangle_*, \nonumber \\
C(t,t') &=& \langle x(t) x(t')\rangle_*, \nonumber \\
G(t,t') &=& \frac{\delta}{\delta \eta(t')}\langle x(t)\rangle_*, \label{eff_param}
\EE
where the average $\langle\cdots\rangle_*$ is performed over realisations of the process in Eq.~\eqref{effective_dynamics}. Eqs.~\eqref{effective_dynamics}-\eqref{eff_param} form a closed system and have to be solved self-consistently.

\subsection{Fixed point analysis}
	
There is no realistic prospect for a general analytical solution of the effective dynamics in Eq.~(\ref{effective_dynamics}). One alternative is to use Monte-Carlo methods to construct sample paths for the effective process and solutions for the dynamic order parameters. For example via the Eissfeller-Opper procedure \cite{eissfeller1992new}, or using the more recent approach in \cite{roy2019numerical}. The latter reference explicitly discusses applications to random Lotka-Volterra systems with Gaussian disorder.

Here we will instead follow \cite{opper1992phase, galla2018dynamically, sidhom2020ecological} and focus on analytical solutions in the parameter regime in which the dynamics approach stable fixed points. This is motivated by observations from the numerical integration of Eq.~\eqref{HLV}. We find that, for certain parameters, the system tends to a unique fixed point, which is independent of initial conditions. Fig.~\ref{fig:cmap} shows examples of parameter regions in which this is the case. Broadly speaking, we observe two different types of behaviour: (i) the population densities converge to a fixed point, or (ii) they diverge. These types of behaviour occur in different regions of parameter space (Fig.~\ref{fig:cmap}). There is a thin boundary between the two regions where other behaviour (e.g. periodic behaviour or persistent irregular motion) can appear, as evidenced by the occasional green or light blue pixel in Fig.~\ref{fig:cmap}. We attribute this to the fact that the system size $N$ is necessarily finite in numerical experiments, and we expect that this behaviour will become increasingly rare as $N\to\infty$.

We will thus assume that each path in the ensemble of trajectories of the effective process eventually arrives at a unique fixed point, $x = \lim_{t\rightarrow\infty} x(t)$. Each realisation of the noise variable $\eta(t)$ in Eq.~(\ref{eff_param}) also approaches a stationary value $\eta$. We note that $x$ and $\eta$ will be random variables, differing across realisations of the effective dynamics. We can then write
\BE
    M &=& \langle x\rangle_*, \nonumber \\
     G(\tau) &=& \lim_{t\rightarrow\infty}G(t+\tau, t), \nonumber \\
    q &=& \lim_{t\rightarrow\infty}C(t+\tau, t) = \langle x^2\rangle_*.\label{corr}
\EE

These relations can be understood as follows: if all realisations of the effective dynamics approach stationary values then $M(t)$ will approach a constant, given by $\avg{x}_*$. Furthermore, we assume that the response function $G(t,t')$ becomes time-translation invariant for large $t$, i.e. $G(t,t')=G(t-t')$. Causality implies that $G(t-t')=0$ for $t<t'$. Finally, given that all trajectories of the effective dynamics approach fixed points, the correlation function $C$ loses all time dependence and so we have written $C(t,t')=q$. This is consistent with Eq.~(\ref{eff_noise}), the noise variables $\eta(t)$ also approach a random but time-independent value for all realisations. The mean of the random variable $\eta$ is zero and using Eq.~(\ref{eff_noise}), its variance is given by
\begin{equation}
    \langle \eta^2\rangle = \alpha q\left[\frac{c}{(1-\chi)^2}+1-c\right], \label{noise} 
\end{equation}
where 
\be\label{eq:chi}
\chi=\int_0^\infty d\tau\,G(\tau).
\ee
From now on, we will write $\eta = \sqrt{q}\,\Sigma z$, where $z$ is a standard Gaussian random variable, and
\be
\Sigma^2=\alpha\left[\frac{ c}{(1-\chi)^2}+1-c\right].
\ee

Setting the time derivative on the left-hand side of Eq.~\eqref{effective_dynamics} to zero, and using Eqs.~\eqref{corr}-\eqref{eq:chi} we find
\BE
x\left[1-ux-\sqrt{q}\Sigma z 
- \alpha x\left(c\frac{\chi}{1-\chi}+\Gamma(1-c)\chi\right)\right]=0. \nonumber \\
\EE

For a given value of $z$ this is to be solved for $x$, subject to the constraint that abundances are non-negative, i.e, $x(z)\geq 0$. Irrespective of the value of $z$, Eq.~\eqref{fxd_point} always has the solution $x(z)=0$. Additionally, a second non-negative solution is possible for some values of $z$. As we will confirm in simulations, the physically meaningful solution is given by
\begin{equation}
x = \mbox{max}\left\{0,\frac{1-\sqrt{q}\Sigma z}{u+\alpha \left(c\frac{\chi}{1-\chi}+\Gamma(1-c)\chi\right)}\right\}.\label{fxd_point}
\end{equation}
For given order parameters $q$ and $\chi$, the function $x(z)$ in Eq.~(\ref{fxd_point}) is therefore piecewise linear, with one piece equal to zero. The denominator in Eq.~\eqref{fxd_point} always comes out positive. Therefore, we have the solution $x(z)>0$ when $z<\Delta$, and $x(z)=0$ when $z\geq \Delta$, with 
 \be\label{eq:delta}
 \Delta=\frac{1}{\Sigma \sqrt{q}}.
 \ee

Given that $z$ is a Gaussian random variable, the abundances of extant species at the fixed point follow a clipped Gaussian distribution. This is similar to what was reported in other random Lotka--Volterra models, see e.g. \cite{bunin2016interaction}. An explicit example of a species abundance distribution can be found for instance in \cite{galla2018dynamically}.
\begin{figure*}[t!]
	\centering
	\includegraphics[width=1\linewidth]{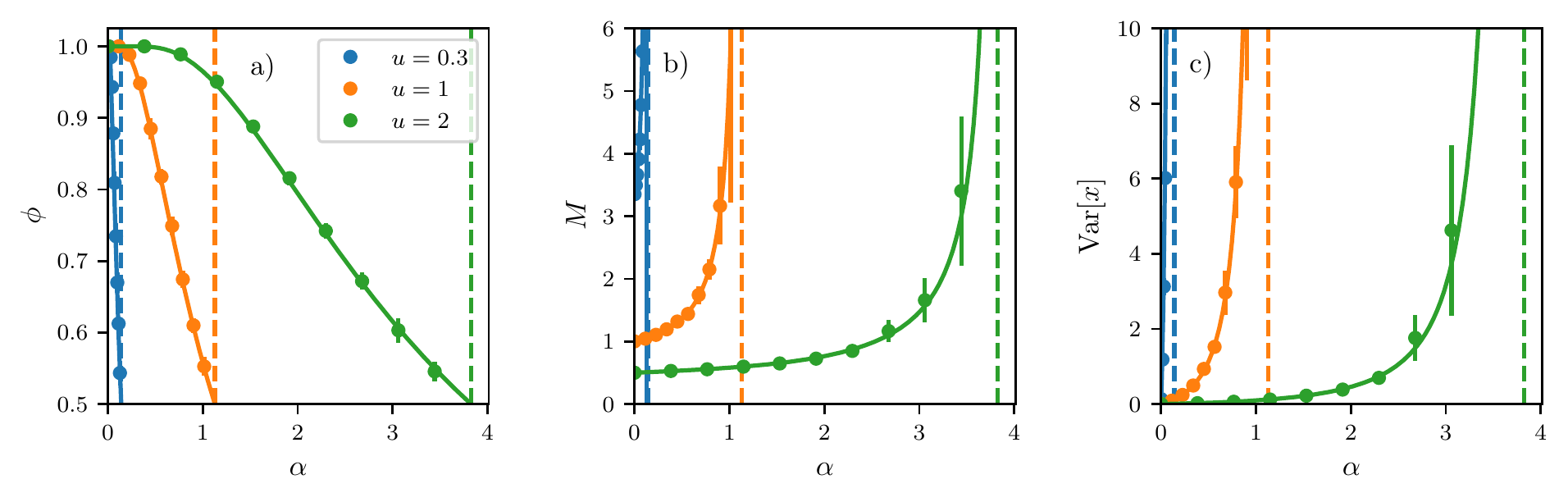}
	\caption{Test of analytical predictions for the order parameters against numerical simulations. The figure shows the fraction of surviving species $\phi$, the mean abundance $M$, and the variance of abundances ($q-M^2$) as a function of the model parameter $\alpha$ (where $P=\alpha cN$ is the number of traits each species in the original system possesses). Lines are from the theory, derived in Eqs.~(\ref{eq:c_less_1}) and (\ref{eq:M_q_alpha}), markers from numerical integration of the gLVE ($N=1000$, $t_{\rm max}=30$, averaged over $10$ realisations of the disorder). The remaining model parameters are $c=0.5$, and $\Gamma=0.3$. Vertical dashed lines indicate the onset of divergence as determined from the theory in Sec.~\ref{sec:stab}.}
	\label{fig:theory_sim}
\end{figure*}

Using this fixed point ansatz, the relations for the order parameters in Eq.~(\ref{corr}) can be written in the following form \cite{bunin2016interaction,galla2018dynamically}
\BE\label{eq:aux}
M&=& \int_{-\infty}^\Delta Dz~x(z),\nonumber \\
\chi&=&\frac{1}{\sqrt{q\Sigma}}\int_{-\infty}^\Delta Dz~\frac{\partial x(z)}{\partial z}, \nonumber \\
q&=& \int_{-\infty}^\Delta Dz~x(z)^2,
\EE
where $Dz = \frac{dz}{\sqrt{2\pi}}e^{-\frac{z^2}{2}}$.
It is now convenient to introduce the following functions
\begin{equation}
f_n(\Delta) := \int_{-\infty}^\Delta Dz\, (\Delta - z)^n, \label{eq:thedefinition}
\end{equation}
for $n=0,1,2$. We then find from Eqs.~(\ref{eq:aux})
\BE
     -\chi\left[u+\alpha \left(c\frac{\chi}{1-\chi}+\Gamma(1-c)\chi\right)\right]&=& f_0(\Delta) , \nonumber\\
     M \frac{u+\alpha \left(c\frac{\chi}{1-\chi}+\Gamma(1-c)\chi\right)}{\sqrt{\alpha q\left(\frac{c}{(1-\chi)^2}+(1-c)\right)}}&=&f_1(\Delta),  \nonumber \\
    \frac{\left[u+\alpha \left(c\frac{\chi}{1-\chi}+\Gamma(1-c)\chi\right)\right]^2}{\alpha \left(\frac{c}{(1-\chi)^2}+(1-c)\right)}&=&f_2(\Delta).\nonumber \\ \label{stat_relations}
\EE
Eqs.~(\ref{stat_relations}) together with Eq.~(\ref{eq:delta}) form a closed set for the  set of unknowns $q,\chi, M$ and $\Delta$, which is to be solved as a function of the model parameters $u, c, \Gamma$ and $
\alpha$.

Recalling that $x(z)>0$ if, and only if, $z<\Delta$ we identify $f_0(\Delta)$ as the fraction of surviving species,
\begin{equation}
	\phi\equiv f_0(\Delta) = \int_{-\infty}^{\Delta} Dz.
\end{equation}

Eqs. \eqref{stat_relations} can be solved parametrically. We fix $u, c, \Gamma$ and $\Delta$ and then solve for the set of $\chi, q, M $ and $\alpha$.

In detail, we find the following cubic equation for $\chi$, valid for $c<1$, 
\begin{align}
0=&f_0 (c (\Gamma -1) f_0-\Gamma  f_0-f_2) \nonumber\\
&+\chi  \left(f_0^2[c+2\Gamma-2 c \Gamma]+ 2 f_0 f_2[1-c] - f_2 u \right)\nonumber \\
&+\chi ^2 (c-1) \left(\Gamma  f_0^2+f_0 f_2-2 f_2 u\right) \nonumber \\
&+ \chi ^3 uf_2  (c-1). \label{eq:c_less_1}
\end{align}
Further, we have from Eqs.~(\ref{stat_relations}),
\BE
 	M &=& \chi\frac{f_1^2}{f_0(f_0-f_2)}, \nonumber \\ 
	q &=& \chi^2 \left(\frac{f_1}{f_2-f_0}\right)^2\frac{f_2}{f_0^2}, \nonumber \\
		\alpha &=& \frac{f_0^2}{f_2} \frac{1}{\chi^2(1-c+\frac{c}{(1-\chi)^2})},
		 \label{eq:M_q_alpha}
\EE
where the $f_n$ are to be evaluated at $\Delta$.

The relations in Eqs.~(\ref{eq:c_less_1}) and (\ref{eq:M_q_alpha}) are also valid for $c=1$, and can then be simplified as outlined in Appendix~\ref{app:c_eq_1}.

The validity of the predictions from Eqs.~(\ref{eq:c_less_1}) and (\ref{eq:M_q_alpha}) is confirmed by direct numerical integration of the gLVE in Fig.~\ref{fig:theory_sim}.

\subsection{Stability analysis} \label{sec:stab}
\subsubsection{Diverging abundance}
{\em Model with $c<1$.} The first and second relations in Eqs.~(\ref{eq:M_q_alpha}) indicate that the order parameters $M$ and $q$ both diverge in the system with $c<1$ when $f_0(\Delta)=f_2(\Delta)$. The latter implies $\Delta =0$. The value of $\alpha$ for which this occurs can (for a given choice of $c, u$ and $\Gamma$) be obtained from the third relation in Eqs.~\eqref{eq:M_q_alpha}, with $\chi$ being the relevant root of Eq.~(\ref{eq:c_less_1}). Using Eq.~(\ref{eq:c_less_1}) the susceptibility $\chi$ is found to remain finite at the transition. We note that $f_0(\Delta)>0$ for all relevant values of $\Delta$.
 
\smallskip

{\em Model with $c=1$.} The fully connected system also shows two types of divergences: (i) The quantities $M$ and $q$ both diverge when $f_0(\Delta)=f_2(\Delta)$, see Eqs.~(\ref{eq:c_eq_1}). The susceptibility then remains finite; (ii) Eqs.~(\ref{eq:c_eq_1}) further indicate, that $M$ and $q$ also diverge in the model with $c=1$ when $f_0^2(\Delta)=u f_2(\Delta)$. This latter condition results in $\alpha=u$. From Eqs.~(\ref{eq:c_eq_1}) the susceptibility $\chi$ is then seen to diverge as well (the divergences of $M$, $q$, and $\chi$ take place simultaneously). 

We note that the divergencies resulting from ${f_0=f_2}$ and ${f_0^2=uf_2}$ can take place at different locations in parameter space for the model with $c=1$. If this is the case, and starting in the stable phase, the divergence that occurs first will determine the loss of stability in the fully connected system. For $u< 1/2$ the transition of type (ii) takes place first as $\alpha$ is increased ($M$, $q$, and $\chi$ diverge), and for $u> 1/2$ the transition of type (i) is instead observed ($q, M$ diverge, $\chi$ remains finite). At present we do not have any further intuition regarding any significance or special role of the value $u=1/2$.

\subsubsection{Linear instability}\label{sec:linear}

The system also shows a linear instability which can be identified using the procedure established in \cite{opper1992phase, galla2005dynamics}. We write $x(t)=x+y(t)$ and $\eta(t)=\sqrt{q}\,\Sigma z +\zeta(t)$, where $y(t)$ and $\zeta(t)$ are small perturbations about the fixed point of the trajectories of the effective process in Eq. \eqref{effective_dynamics}. Expanding to first order in these perturbations we find that
\begin{align}
	\dot{y}(t)=&x\left(-uy(t)-\alpha\int_{-\infty}^tdt'\, K\left(t,t'\right)y\left(t'\right)- \zeta(t) \right) \nonumber \\
	&+ y(t)\left(1-ux-\alpha x\int_{-\infty}^tdt'\, K\left(t,t'\right) - \sqrt{q}\,\Sigma z \right),\label{perturbation}
\end{align}
with $K(t,t')=[cG(\mathbb{I}-G)^{-1}
+\Gamma(1-c)G](t,t')$. We also have the self-consistency relation
\begin{align}
	\left\langle  \zeta(t) \zeta\left(t'\right) \right\rangle = \alpha\big[ c&(\mathbb{I}-G)^{-1}D(\mathbb{I}-G^T)^{-1}\nonumber \\
	&+(1-c)D \big]\left(t,t'\right), \label{sc}
\end{align}
where $D\left(t,t'\right) = \left\langle y(t)y\left(t'\right)\right\rangle_*$.

When $x=0$, Eq.~\eqref{perturbation} becomes
\begin{equation}
	\dot{y}(t) = y(t) \left(1 - \sqrt{q}\,\Sigma z\right). \label{member}
\end{equation}
Eq.~(\ref{fxd_point}), together with the observation that the denominator in this equation is strictly positive, implies that $1 - \sqrt{q}\,\Sigma z<0$ when $x=0$. This allows us to conclude that perturbations on extinct species decay, and do not contribute to any linear instability. 

For fixed points $x>0$ we find from Eqs.~\eqref{fxd_point} and \eqref{perturbation} that 
\begin{equation}
	\dot{y}(t) = -x\left(uy(t) + \alpha\int_{-\infty}^tdt'\, K\left(t-t'\right)y\left(t'\right) + \zeta(t)\right). \label{lineq}
\end{equation}
To identify the onset of linear instability we follow  \cite{opper1992phase,galla2005dynamics}. We move to Fourier space, writing $\omega$ for the variable conjugate to time $t$, and using tildes to indicate Fourier transforms.

Focusing on the mode with $\omega=0$ and following steps similar to those in \cite{opper1992phase,galla2005dynamics,galla2018dynamically} we then find from Eq.~\eqref{lineq}
\begin{align}
\avg{|\tilde y(0)|^2}=\Biggr[\phi^{-1}&\left(u+\alpha c\frac{\chi}{1-\chi}+\alpha\Gamma(1-c)\chi\right)^2 \nonumber\\
&-\alpha\left(\frac{c}{(1-\chi)^2}+1-c\right)\Biggr]^{-1}.
\end{align}
The left-hand side is manifestly non-negative, so a change of sign of the expression inside the square bracket on the right-hand side indicates an inconsistency (and divergence of $\avg{|\tilde y(0)|^2}$).
Using Eqs.~\eqref{stat_relations} this is shown to occur when 
\begin{equation}
\alpha\left[\frac{c}{(1-\chi)^2}+1-c\right](f_0-f_2)=0.
\end{equation}

For $c<1$ the expression in the square brackets is never zero. This leaves us with the condition $f_0=f_2$, which is the same as we obtained for the divergence of $M$ and $q$. If $c=1$, the term in the square bracket is zero if $\chi\rightarrow\infty$, which using Eq.~\eqref{eq:c_eq_1} we can write as $f_0^2-uf_2=0$.

From this, we conclude that in our model the linear instability is always accompanied by the instability with diverging mean abundance. This is markedly different from the behaviour of the gLVE model with Gaussian random interactions. In this Gaussian model there are instances where the linear instability sets in as the variance of interactions is increased, but where abundances remain finite and the divergence only occurs at a later point at even higher variance of the interactions. This leads to a phase with multiple attractors between the two transitions \cite{bunin2016interaction,bunin2017ecological,biroli2018marginally}. Our analysis indicates that the model with Hopfield-like couplings does not have such a multiple-attractor regime.

\section{Phase diagram and further behaviour of the model}\label{sec:pg}

\subsection{Phase diagram for the fully connected system ($c=1$)}

\begin{figure*}[t!]
	\centering
	\includegraphics[width=1\linewidth]{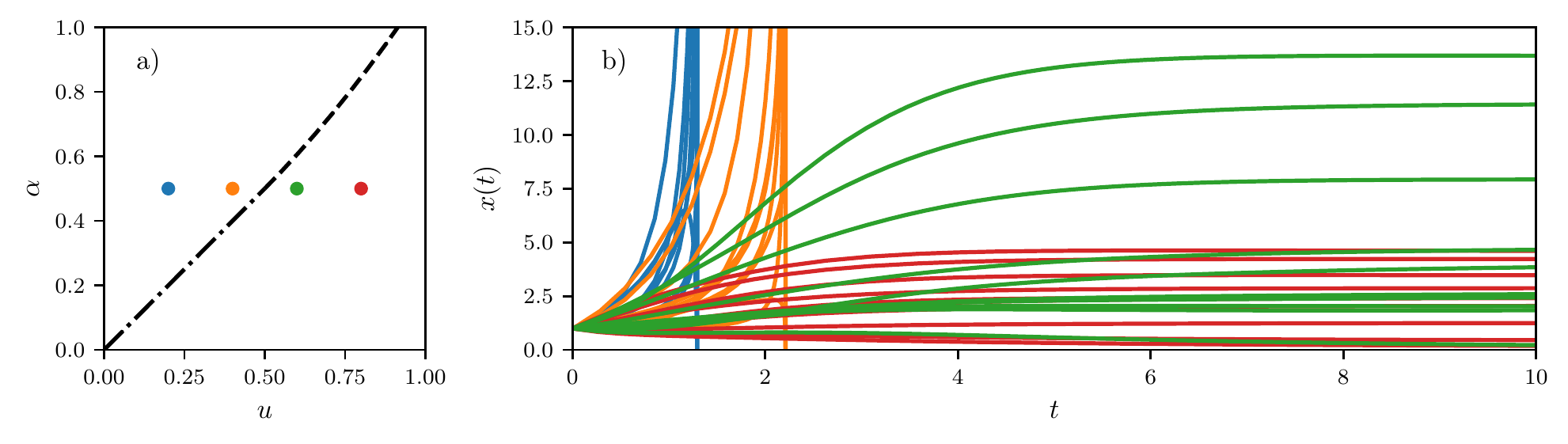}
	\caption{Phase behaviour of the fully connected model ($c=1$). Panel (a): Phase diagram for the model with $c=1$, the only model parameters are then $u$ and $\alpha$. The system is stable to the right of the lines. At the dot-dashed line ($u<1/2$) $q, M$ and $\chi$ all diverge, and at the dashed line $M$ and $q$ diverge, but $\chi$ remains finite. Panel (b): Illustration of the behaviour of the abundances of individual species in the two different phases (convergence to a fixed point shown in green and red, diverging abundances in orange and blue).}
	\label{fig:pg_c_eq_1}
\end{figure*}

The phase diagram of the fully connected model is shown in  Fig.~\ref{fig:pg_c_eq_1}(a). We recall that, for $c=1$, the only model parameters are the self-interaction coefficient $u$ and the ratio of the number of traits to the number of interspecies interactions in the original pool ($\alpha=P/cN$). For a fixed value of $\alpha$, the system shows a unique stable fixed point for $u>u_c(\alpha)$, where $u_c(\alpha)$ marks the onset of instability. The line in Fig.~\ref{fig:pg_c_eq_1}(a), obtained from Eq.~\eqref{eq:c_eq_1}, shows the phase boundary between the stable and unstable regions. At this boundary $M$ and $q$ diverge, and if $u_c<1/2$ we also observe a divergence of $\chi$.

The two types of trajectory in the stable and divergent phases are illustrated in the right panel of Fig.~\ref{fig:pg_c_eq_1}(b). In the stable phase the system reaches a fixed point, for any one realisation of the interaction matrix (two examples are illustrated in green and red respectively).

Fig.~\ref{fig:pg_c_eq_1}(b) also shows  two examples in which the species abundances diverge (blue and orange). The divergence  occurs at a finite time. We will discuss this further in Sec.~\ref{sec:div_finite_t}.

\subsection{Phase diagram for connectivity $c<1$}

\begin{figure*}[t]
	\centering	\includegraphics[width=1\linewidth]{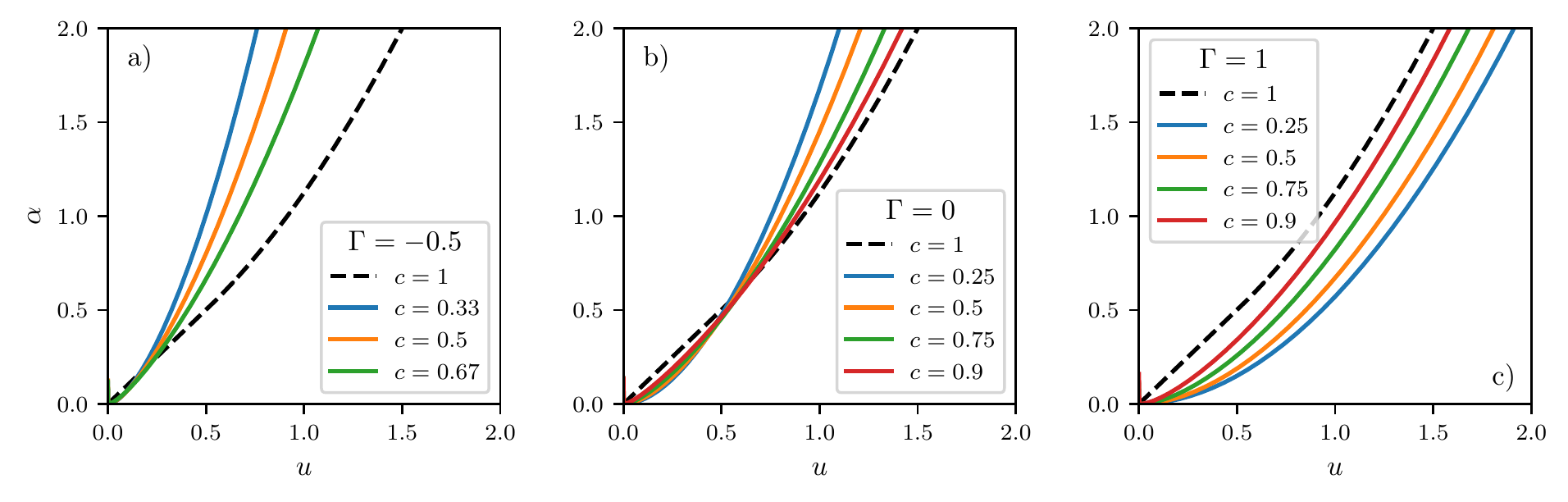}
	\caption{Phase diagram for different choices of the connectivity $c$, and the symmetry parameter $\Gamma$. The coloured lines in each panel indicate where the linear instability occurs.  The instability coincides with the divergence of $M$ and $q$. The system is stable to the right of the line, abundances diverge on the left.}
	\label{fig:pg_c_general}
\end{figure*}

Fig.~\ref{fig:pg_c_general} shows how the phase diagram for the system with $c<1$ depends on the connectivity $c$ and the symmetry parameter $\Gamma$. In all cases there is a single phase boundary, where the divergence of $M$ and $q$ and the onset of linear instabilities coincide. This phase boundary separates a region where trajectories converge to a single globally stable fixed point (phase to the right of the line), from a region where trajectories are unbounded and diverge in finite time (phase to the left).

The phase diagrams in Figs.~\ref{fig:pg_c_eq_1} and \ref{fig:pg_c_general} show that the system is in the stable phase for small values of $\alpha$ (i.e. a small number of traits relative to the number of species in the initial pool), or large values of $u$ (i.e.  large negative self-interaction). This is the consequence of two competing effects, the self-interaction (parametrised by $u$) which stabilises the system, and the interaction between species (induced by competition of similar species) which promotes instability. When $u$ is large and/or $\alpha$ is small, the stabilizing effect of the intra-species interaction dominates over the interactions across species. In the extreme limit $\alpha=0$ (no interaction between different species), each abundance follows a separate logistic equation, $\dot x_i = x_i (1-ux_i)$, and converges to $x_i=1/u$. When $\alpha$ is small but non-zero, the system consists of weakly interacting species. The effect of the interactions between species is then a small perturbation to the logistic behaviour of individual species, and does not change the convergence to a fixed point. This can be confirmed from Eqs. \eqref{eq:c_less_1} and \eqref{eq:M_q_alpha} by taking the limit $\alpha\rightarrow0$, which results in all species surviving with fixed point abundance $x_i^*=1/u$  ($\phi\rightarrow1$, $M\rightarrow \frac{1}{u}$, and $\text{Var}[x]\rightarrow 0$). A similar result is obtained for $u\gg 1$ at fixed value of $\alpha$.

Conversely, for low values of $u$ or large values of $\alpha$ the system is unstable. In this situation, the stabilising self-interaction is not sufficient to overcome the destabilising effect of the random interactions between species.

The most interesting behaviour takes place at the phase boundary, where the effect of the intraspecific and interspecific nonlinearities are of comparable magnitude. From Eqs. \eqref{eq:c_less_1} and \eqref{eq:M_q_alpha} we can conclude  that $\phi\rightarrow 1/2$, and $M\sim(\alpha_c-\alpha)^{-1}$ as the system approaches the instability (from the stable phase). Further details can be found in Appendix~\ref{app:limit}.

We further note that decreasing the value of the symmetry parameter $\Gamma$, increases the range of the stable region in the phase diagrams in Fig.~\ref{fig:pg_c_general}. This is similar to the effect of increasing the fraction of predator-prey interactions in Lotka--Volterra models with Gaussian interactions \cite{allesina2012stability,bunin2016interaction,galla2018dynamically}. Indeed, the effect of a reduction of $\Gamma$ is to increase the fraction of species pairs $i,j$ with $c_{ij}=1$ and $c_{ji}=0$, that is the proportion of uni-directional interactions.

\begin{figure*}[t!!]
	\centering
    \includegraphics[width=1\linewidth]{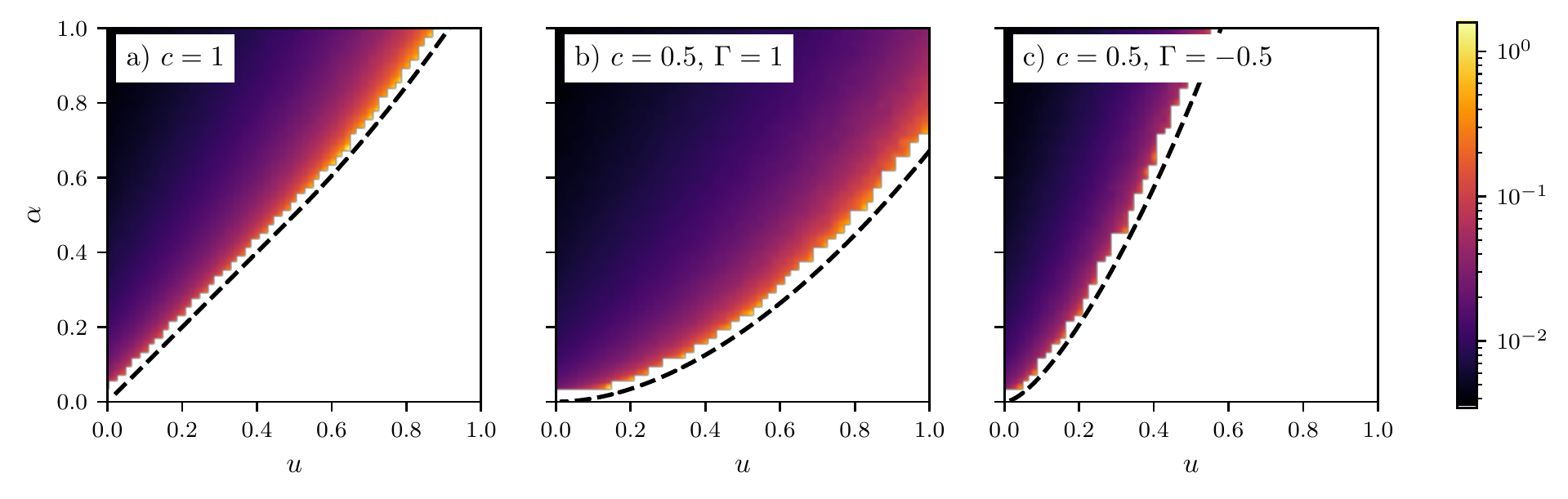} 
	\caption{Finite-time divergence of abundances. The heatmaps indicate the time, $t_{\rm div}$, at which abundances diverge, for initial conditions $x_i(0)=1$. Data is obtained from numerical integration of the gLVE. The dotted line is the phase boundary predicted by the theory. To the right of the phase boundary the system is in the stable phase, so that no divergence occurs.}
	\label{fig:t_div_1}
\end{figure*}

Interestingly, the effect of varying the `connectance' $c$ is not straightforward. As can be seen in Fig.~\ref{fig:pg_c_general} an increased connectivity can, depending on the other model parameters, turn a previously stable system into an unstable one, or vice versa, stabilise a previously unstable system.

\subsection{Finite-time divergence of the mean abundance}\label{sec:div_finite_t}

As mentioned earlier, the divergence of the abundances in the divergent phase occurs at finite time. This has previously been reported in the model with Gaussian interactions \cite{roy2019numerical}, and can be justified heuristically from the Lotka-Volterra equations. Indeed, Eq.~\eqref{HLV} has a second-order non-linearity in the abundances $x_i$. This can lead to dynamics of the form $\dot{x} \sim x^2$, which in turn implies a solution of the form $x(t)=(c-t)^{-1}$, where $c$ is an integration constant. This results in a  divergence at finite time.

\begin{figure}[h]
	\centering
	\includegraphics[width=1\linewidth]{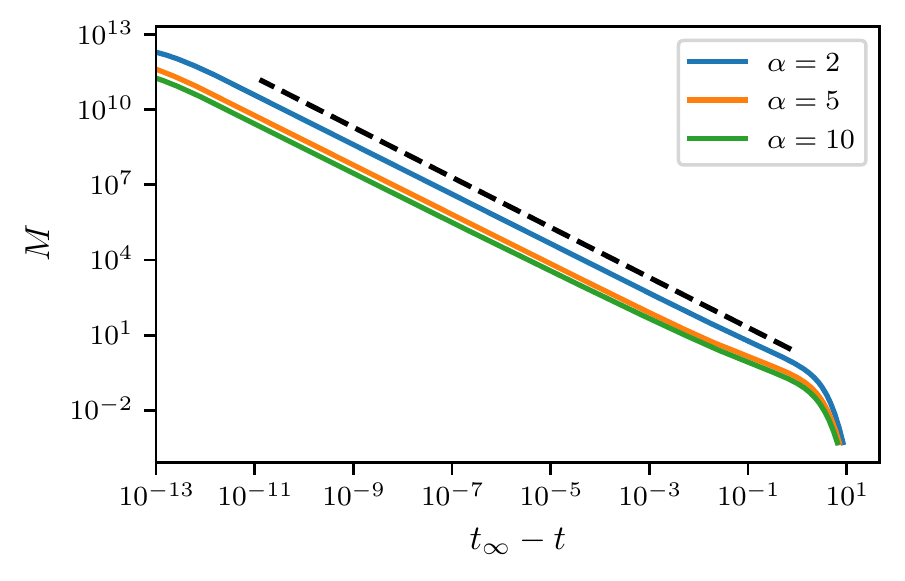}
	\caption{Divergence of $M$ for initial conditions uniformly distributed in $(0, 10^{-3})$ and parameters $N=1000$, $u=1$, and different values of $\alpha$. The dotted black line corresponds to $(t_\infty-t)^{-1}$. The deviation from $M\sim (t_\infty-t)^{-1}$ close to the divergence is attributed to numerical error.}

	\label{fig:t_div_2}
\end{figure}

Fig.~\ref{fig:t_div_1} shows the time, $t_{\rm div}$, at which the divergence occurs for different choices of the model parameters. This time grows as one approaches the stability line (from inside the unstable phase). When the stability line is crossed (into the stable phase), the time-to-divergence diverges itself ($t_{\rm div}\to\infty$), i.e. the divergence no longer occurs. Results from the numerical integration of the gLVE suggest that the divergence of the abundances is of the form $M\sim(t_{\rm div}-t)^{\nu}$, where  $\nu\approx 1$, as shown in Fig.~\ref{fig:t_div_2}. This behaviour appears to be independent of initial conditions, the values of the parameters $u$ and $\alpha$, and the initial number of species $N$.

\section{Reduced interaction matrix and its eigenvalue spectrum}\label{sec:reduced_matrix}

Ref.~\cite{baron2022non} recently established a close connection between different instabilities in the Gaussian random Lotka--Volterra model and the eigenvalue spectrum of the interaction matrix of the surviving species. More specifically, the spectrum of this reduced interaction matrix is composed of a bulk region and a potential outlier eigenvalue. As parameters are changed (starting from within the stable phase) either the bulk spectrum or the outlier eigenvalue can cross into the right half of the complex plane. In the Gaussian model, the crossing of the outlier is associated with a transition marked by the divergence of abundances, and a crossing of the bulk with a linear instability.

In this section, we explore in numerical simulations how the different transitions in the gLVE model with Hopfield-like interactions relate to the eigenvalue spectrum of the matrix of interactions between surviving species.

\subsection{Spectrum of the original interaction matrix}
\begin{figure}[t!]
	\centering
\includegraphics[width=1\linewidth]{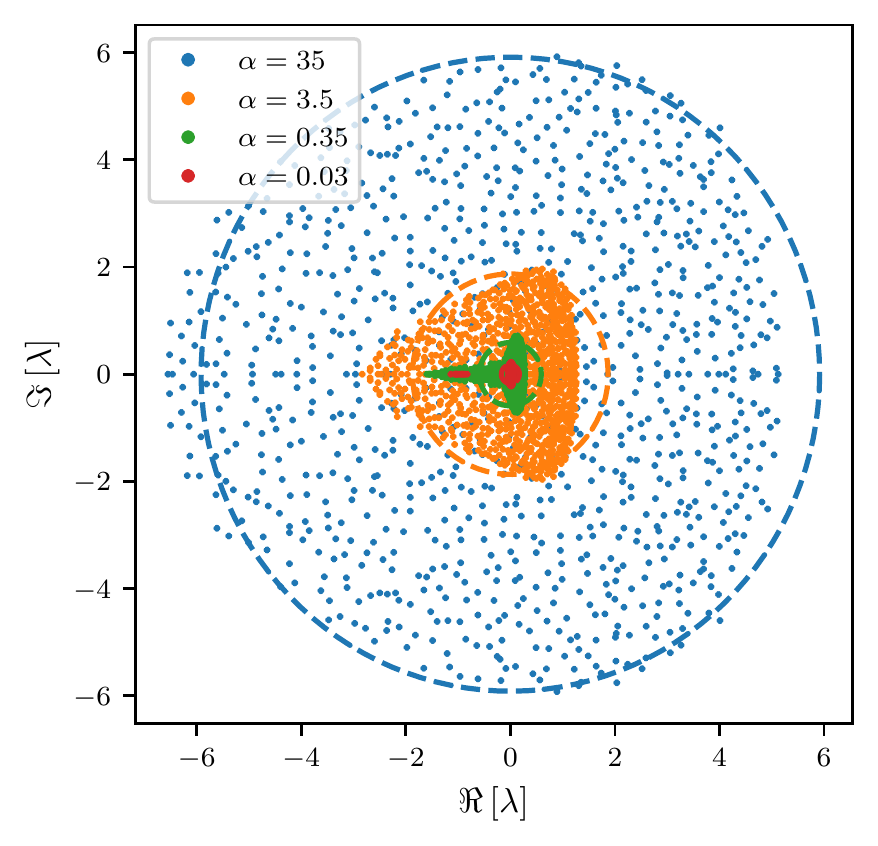}
	\caption{Examples of the eigenvalue spectrum of the original interaction matrix. The dashed lines are the na\"ive predictions of Eq.~\eqref{gaussian_spectrum}. Model parameters are $c=0.4,\Gamma=c/(c-1)$ and $N=5000$.
	\label{fig:eig}}
\end{figure}
Before we discuss the spectra of the reduced interaction matrix, we make a few remarks on the initial interaction matrix $\alpha_{ij}=c_{ij}J_{ij}$ among all species. Throughout this section we set the diagonal elements of this matrix to zero, the only effect of self-interaction (the term $-ux_i$) is a simple shift of this spectrum. In the large-$N$ limit the central limit theorem applies to $J_{ij}=-\frac{1}{cN}\sum_{\mu=1}^{\alpha c N}\xi_i^\mu\xi_j^\mu$, so each off-diagonal entry $\alpha_{ij}$ of the interaction matrix is either a Gaussian random variable (if $c_{ij}=1$) or equals zero (if $c_{ij}=0$). The variance of $\alpha_{ij}$ is
\be
\mbox{Var}(\alpha_{ij}) = \frac{1}{cN^2}\sum_{\mu, \mu'}^{\alpha c N}\left\langle\xi_i^{\mu}\xi_i^{\mu'}\xi_j^{\mu}\xi_j^{\mu'}\right\rangle  = \frac{\alpha}{N}.
\ee

Calculating the correlations between pairs of elements we obtain
\begin{align}
\nonumber\text{Corr}&[\alpha_{ij},\alpha_{nm}]
=\frac{\langle \alpha_{ij}\alpha_{nm}\rangle-\langle \alpha_{ij}\rangle\langle \alpha_{nm}\rangle}{\mbox{Var}(\alpha_{ij})} \\
&= 
\begin{cases}
\Gamma (1-c) + c & (i,j)=(m,n) \\
1 & (i,j)=(n,m) \\
0 & \text{else}
\end{cases} \ ,\label{crosscorr}
\end{align}
where we have used Eq. \eqref{prop_c} and the fact that $J_{ij}$ is symmetric. This means that only diagonally opposed pairs of elements are correlated, and that their correlation is determined by both, $\Gamma$ and $c$.

Based on a theory that only takes into account correlations between diagonally opposed matrix entries, one might then expect an elliptic spectrum \cite{sommers1988spectrum}, with support given by the ellipse
\begin{equation}
	\left(\frac{x}{\sqrt{\alpha}(1+\tau)}\right)^2 + \left(\frac{y}{\sqrt{\alpha}(1-\tau)}\right)^2 = 1, \label{gaussian_spectrum}
\end{equation}
with $\tau =\Gamma(1-c)+c$. However, as illustrated in Fig.~\ref{fig:eig}, this is an approximation to the true spectrum at best for large values of $\alpha$. For intermediate values of $\alpha$ (an example is shown in orange in the figure), the eigenvalue spectrum appears to have a triangular shape, and for small values of $\alpha$  (shown in green), the spectrum becomes even more skewed and eventually appears to consist of two separate components (example shown in red). While we cannot fully exclude finite-size effects (the spectra in Fig.~\ref{fig:eig} are for $N=5000$), we believe that the deviations from an elliptical spectrum in Eq.~(\ref{gaussian_spectrum}) are due to  higher-order correlations between entries of the interaction matrix. For example, it has been shown in Ref.~\cite{aceituno2019universal} that cyclic correlations can result in eigenvalue spectra with shapes similar to the ones in Fig.~\ref{fig:eig}. For $c=1$ the interaction matrix is a (scaled and shifted) Wishart matrix, so its spectrum follows the Marcenko-Pastur law.

\subsection{Eigenvalues of the reduced interaction matrix}

\begin{figure}[t]
	\centering
	\includegraphics[width=1\linewidth]{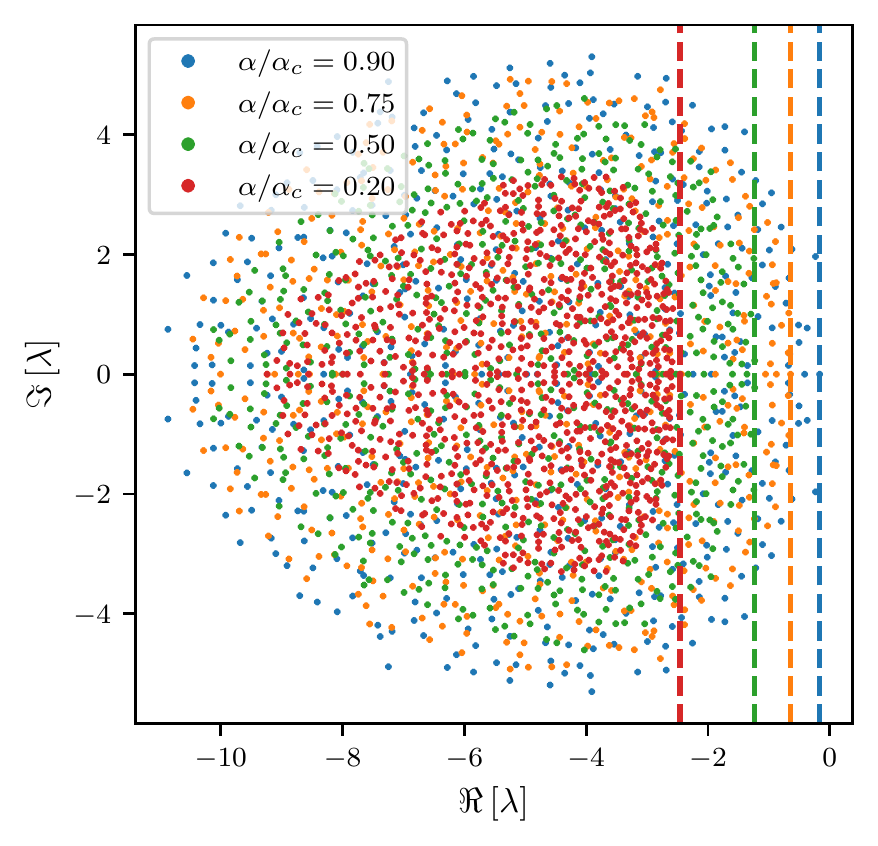}
		\caption{Eigenvalue spectrum of the reduced interaction matrix for $u=5$, $c=0.5$, $\Gamma=c/(c-1)$, for different choices of the model parameter $\alpha$. The vertical dashed lines indicate the real part of the right-most eigenvalue.}
	\label{fig:eig2}
\end{figure}

\begin{figure*}[t]
	\centering
    \includegraphics[width=1\linewidth]{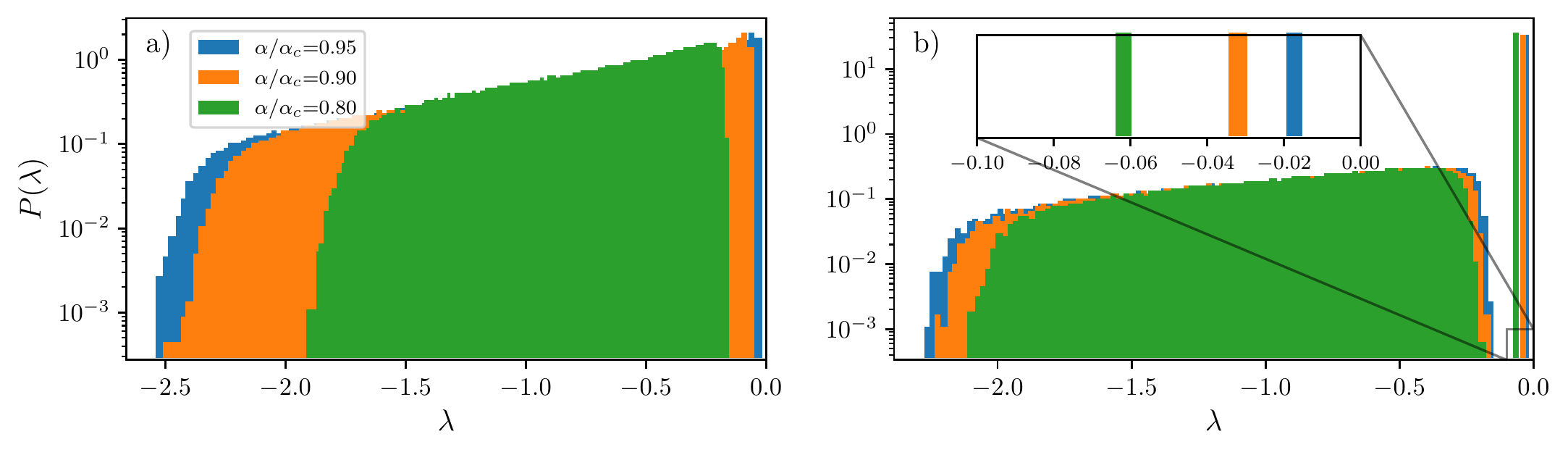}
	\caption{Eigenvalue spectrum of the reduced interaction matrix in the fully connected system. The matrices are symmetric, and their eigenvalues are therefore real-valued. Panel (a) is for $u=0.7$, panel (b) for $u=0.3$.}
	\label{fig:eig3}
\end{figure*}

We now conclude the analysis of the model with a numerical study of the spectra of the reduced interaction matrix, that is, the interaction matrix between species that survive at the fixed points of the gLVE. 

Fig.~\ref{fig:eig2} shows the spectra of this matrix for the case $c<1$, and for a choice of $\Gamma$ less than one. This means that the initial interaction matrix is not symmetric. The reduced matrix is not symmetric either, and as a consequence its eigenvalues will generally be complex. As seen in Fig.~\ref{fig:eig2}, the spectrum is not elliptic, and we have found no evidence of an outlier eigenvalue in this scenario ($c<1$). In the figure we have fixed $u$, and varied $\alpha$. The data suggests that the phase transition at $\alpha=\alpha_c(u)$ coincides with the point at which the right-most bulk eigenvalue crosses the imaginary axis into the right half-plane.
 
In Fig.~\ref{fig:eig3} we study the fully connected system for two different values of the self-interaction strength $u$. The original interaction matrix in the fully connected model is symmetric by construction, and so is the reduced interaction matrix. As a consequence all eigenvalues are real. 

Panel (a) focuses on the case $u>1/2$. We find no signs of outlier eigenvalues, and again the data indicates that the transition to instability occurs when the leading bulk eigenvalue crosses into the positive half of the real axis.

Panel (b) shows a scenario in which  $u<1/2$. In contrast with the situation in (a), an outlier eigenvalue now becomes apparent, and the transition to instability in the gLVE at $\alpha=\alpha_c(u)$ now appears to coincide with the point at which the outlier becomes positive.

\begin{table}[b]
\centering
\begin{tabular}{llcc}
\toprule
~~~$c<1$~~~&  & $q,M$ diverge & bulk spectrum \\
       &   & $\chi$ remains finite & crosses axis \\[0.5em]
       \hline  \\ [-1.5ex]
~~$c=1$~~    & ~~$u>1/2$~~ & $q, M$ diverge  & bulk spectrum  \\
        &   & $\chi$ remains finite & crosses axis \\[0.75em]
& $~~u<1/2~~$  & $q, M, \chi$ & outlier eigenvalue \\
&  &  all diverge & crosses axis \\
&  &  (at $u=\alpha$) &   \\
\hline\hline
\end{tabular}
\caption{Types of phase transition in the gLVE model with Hopfield-like interactions. The table summarises the different transitions, giving details about the nature of the divergence at the transition, and the associated behaviour of the spectrum of the reduced interaction matrix.}\label{tab:pg_table}
\end{table}

The connection between the different types of transition and the behaviour of the spectrum of the reduced interaction matrix is shown in Table~\ref{tab:pg_table}. We recall that the mean abundance and the second moment of the abundances diverge at all transitions, and that the onset of the linear instability always coincides with the point of diverging abundances. There are thus only two types of transition, one in which the susceptibility remains finite ($\chi<\infty$), and another for which it diverges ($\chi\to\infty$). The table indicates that the former transition ($\chi$ finite) appears to coincide with the bulk spectrum of the reduced matrix crossing into the right half of the complex plane. The transition at which $\chi\to \infty$ (along with the divergences of $M$ and $q$) on the other hand seems to be seen when the outlier eigenvalue of the reduced matrix in the fully connected system reaches the origin.

We stress that these are numerical observations, and that these findings should therefore be seen mostly as conjectures at this point.  In principle, the spectrum of the reduced interaction matrix can likely be calculated in our model, adapting the method used in Ref.~\cite{baron2022non}. However, this involves a substantial calculation and is beyond the scope of the current paper.

\section{Discussion}\label{sec:concl}

To summarise, we have carried out a generating functional analysis of a random generalised Lotka--Volterra system with interactions determined by niche overlap. Species interactions in the model are governed by Hopfield-like couplings subject to mild dilution (the remaining connectivity is still extensive). We have computed the statistics of surviving species in the stable fixed point phase, and we have analytically determined the onset of instability. Similar to the gLVE with Gaussian interactions, asymmetry in the connectivity matrix promotes stability. That is to say, the system becomes more stable when there is a larger fraction of unidirectional interactions ($c_{ij}=1$, but $c_{ji}=0$). In contrast with the Gaussian model, the linear instability against small perturbations cannot be separated from an instability at which species abundances diverge. As a consequence, there is no phase with multiple stable fixed points our model. Despite some common features, the statistical mechanics of the Gaussian and Hopfield-like models are therefore rather distinct.

Our analysis shows two types of transitions to divergent abundances, one in which the integrated response $\chi$ remains finite, and another in which $\chi$ diverges. This raises interesting questions about the exact nature of memory onset in the system (a diverging integrated response indicates persistent memory of perturbations). Future work could focus on the precise shape of the response function, where the numerical methods in \cite{roy2019numerical} might prove particularly useful. Given that the fully connected system has symmetric couplings it would also be interesting to see how crossing each of the different types of transition affects the energy landscape. A natural approach here might be the replica method and suitable levels of replica symmetry breaking \cite{biroli2018marginally,Altieri2022}.

Numerical simulations provide evidence that the transition at which the integrated response remains finite ($\chi<\infty$) is associated with the bulk spectrum of the reduced interaction matrix (the matrix of interactions between extant species) crossing the axis. The transition at which $\chi$ diverges on the other hand appears to be signalled by an outlier eigenvalue crossing the imaginary axis. 

These findings in simulations reinforce the intriguing analytical result obtained recently in \cite{baron2022non}. Namely, the eigenvalues of the interaction matrix in the community of surviving species can be used to decide the stability of {\em feasible} equilibria, that is to say, fixed points with non-negative species abundances. In the traditional approach to ecosystem stability by Robert May \cite{may1972will}, based on the eigenvalue spectra of random matrices, no actual dynamics are specified, and the feasibility of the assumed equilibria remains unclear. Any fixed point of the generalised Lotka--Volterra model on the contrary is feasible by construction. The study of the spectra of reduced interaction matrices resulting from Lotka--Volterra dynamics can therefore contribute to establishing how May's approach can be adapted to include feasible equilibria. Noting that previous work \cite{bunin2016interaction, baron2022non} has shown that the statistics of the reduced interaction matrix in random Lotka-Volterra models can be quite different from those of the original interaction matrix, it would be interesting to study the statistics of the $J_{ij}$ among survivors in the present model in more detail. In particular, the niche overlap between surviving species.

On a broader level, our study highlights two common facets of work on the statistical physics of complex systems, which were also seen for example 30-40 years ago when physicists studied neural networks, or 15-20 years ago when a number of physicists worked on the Minority Game. On the one hand, tools from physics can make a difference in problems from other disciplines. In our system (and other models of complex ecosystems more generally) this is the study of feasible equilibria with methods from spin glass physics. At the same time, studying problems arising in other areas can reveal new types of physics and complexity, which one would perhaps not find within the strict boundaries of traditional physics. In our case, these are the different types of phase transition in the generalised Lotka--Volterra model. We think that this mutually beneficial relation of physics and adjacent disciplines is what makes the field of complex systems particularly attractive.

\section*{Acknowledgements}

Partial financial support has been received from the Agencia Estatal de Investigaci\'on and Fondo Europeo de Desarrollo Regional (FEDER, UE) under project APASOS (PID2021-122256NB-C21/PID2021-122256NB-C22), and the Maria de Maeztu program for Units of Excellence, CEX2021-001164-M funded by  MCIN/AEI/10.13039/501100011033.
 
\bibliography{refs.bib}

\begin{thebibliography}{42}%
\makeatletter
\providecommand \@ifxundefined [1]{%
 \@ifx{#1\undefined}
}%
\providecommand \@ifnum [1]{%
 \ifnum #1\expandafter \@firstoftwo
 \else \expandafter \@secondoftwo
 \fi
}%
\providecommand \@ifx [1]{%
 \ifx #1\expandafter \@firstoftwo
 \else \expandafter \@secondoftwo
 \fi
}%
\providecommand \natexlab [1]{#1}%
\providecommand \enquote  [1]{``#1''}%
\providecommand \bibnamefont  [1]{#1}%
\providecommand \bibfnamefont [1]{#1}%
\providecommand \citenamefont [1]{#1}%
\providecommand \href@noop [0]{\@secondoftwo}%
\providecommand \href [0]{\begingroup \@sanitize@url \@href}%
\providecommand \@href[1]{\@@startlink{#1}\@@href}%
\providecommand \@@href[1]{\endgroup#1\@@endlink}%
\providecommand \@sanitize@url [0]{\catcode `\\12\catcode `\$12\catcode `\&12\catcode `\#12\catcode `\^12\catcode `\_12\catcode `\%12\relax}%
\providecommand \@@startlink[1]{}%
\providecommand \@@endlink[0]{}%
\providecommand \url  [0]{\begingroup\@sanitize@url \@url }%
\providecommand \@url [1]{\endgroup\@href {#1}{\urlprefix }}%
\providecommand \urlprefix  [0]{URL }%
\providecommand \Eprint [0]{\href }%
\providecommand \doibase [0]{http://dx.doi.org/}%
\providecommand \selectlanguage [0]{\@gobble}%
\providecommand \bibinfo  [0]{\@secondoftwo}%
\providecommand \bibfield  [0]{\@secondoftwo}%
\providecommand \translation [1]{[#1]}%
\providecommand \BibitemOpen [0]{}%
\providecommand \bibitemStop [0]{}%
\providecommand \bibitemNoStop [0]{.\EOS\space}%
\providecommand \EOS [0]{\spacefactor3000\relax}%
\providecommand \BibitemShut  [1]{\csname bibitem#1\endcsname}%
\let\auto@bib@innerbib\@empty
\bibitem [{\citenamefont {Edwards}\ and\ \citenamefont {Anderson}(1975)}]{edwards1975theory}%
  \BibitemOpen
  \bibfield  {author} {\bibinfo {author} {\bibfnamefont {S.~F.}\ \bibnamefont {Edwards}}\ and\ \bibinfo {author} {\bibfnamefont {P.~W.}\ \bibnamefont {Anderson}},\ }\href@noop {} {\bibfield  {journal} {\bibinfo  {journal} {Journal of Physics F: Metal Physics}\ }\textbf {\bibinfo {volume} {5}},\ \bibinfo {pages} {965} (\bibinfo {year} {1975})}\BibitemShut {NoStop}%
\bibitem [{\citenamefont {M{\'e}zard}\ \emph {et~al.}(1987)\citenamefont {M{\'e}zard}, \citenamefont {Parisi},\ and\ \citenamefont {Virasoro}}]{mezard1987spin}%
  \BibitemOpen
  \bibfield  {author} {\bibinfo {author} {\bibfnamefont {M.}~\bibnamefont {M{\'e}zard}}, \bibinfo {author} {\bibfnamefont {G.}~\bibnamefont {Parisi}}, \ and\ \bibinfo {author} {\bibfnamefont {M.~A.}\ \bibnamefont {Virasoro}},\ }\href@noop {} {\emph {\bibinfo {title} {Spin glass theory and beyond: An Introduction to the Replica Method and Its Applications}}},\ Vol.~\bibinfo {volume} {9}\ (\bibinfo  {publisher} {World Scientific Publishing Company},\ \bibinfo {year} {1987})\BibitemShut {NoStop}%
\bibitem [{\citenamefont {Coolen}(2001{\natexlab{a}})}]{coolen_stat}%
  \BibitemOpen
  \bibfield  {author} {\bibinfo {author} {\bibfnamefont {A.~C.~C.}\ \bibnamefont {Coolen}},\ }in\ \href@noop {} {\emph {\bibinfo {booktitle} {Neuro-Informatics and Neural Modelling}}},\ \bibinfo {series} {Handbook of Biological Physics}, Vol.~\bibinfo {volume} {4},\ \bibinfo {editor} {edited by\ \bibinfo {editor} {\bibfnamefont {F.}~\bibnamefont {Moss}}\ and\ \bibinfo {editor} {\bibfnamefont {S.}~\bibnamefont {Gielen}}}\ (\bibinfo  {publisher} {North-Holland},\ \bibinfo {year} {2001})\ pp.\ \bibinfo {pages} {553--618}\BibitemShut {NoStop}%
\bibitem [{\citenamefont {Coolen}(2001{\natexlab{b}})}]{Coolen_dyn}%
  \BibitemOpen
  \bibfield  {author} {\bibinfo {author} {\bibfnamefont {A.~C.~C.}\ \bibnamefont {Coolen}},\ }in\ \href@noop {} {\emph {\bibinfo {booktitle} {Neuro-Informatics and Neural Modelling}}},\ \bibinfo {series} {Handbook of Biological Physics}, Vol.~\bibinfo {volume} {4},\ \bibinfo {editor} {edited by\ \bibinfo {editor} {\bibfnamefont {F.}~\bibnamefont {Moss}}\ and\ \bibinfo {editor} {\bibfnamefont {S.}~\bibnamefont {Gielen}}}\ (\bibinfo  {publisher} {North-Holland},\ \bibinfo {year} {2001})\ pp.\ \bibinfo {pages} {619--684}\BibitemShut {NoStop}%
\bibitem [{\citenamefont {Coolen}\ \emph {et~al.}(2005)\citenamefont {Coolen}, \citenamefont {Kuehn},\ and\ \citenamefont {Sollich}}]{coolen_kuehn_sollich}%
  \BibitemOpen
  \bibfield  {author} {\bibinfo {author} {\bibfnamefont {A.~C.~C.}\ \bibnamefont {Coolen}}, \bibinfo {author} {\bibfnamefont {R.}~\bibnamefont {Kuehn}}, \ and\ \bibinfo {author} {\bibfnamefont {P.}~\bibnamefont {Sollich}},\ }\href@noop {} {\emph {\bibinfo {title} {Theory of Neural Information Processing Systems}}}\ (\bibinfo  {publisher} {Oxford University Press, Oxford UK},\ \bibinfo {year} {2005})\BibitemShut {NoStop}%
\bibitem [{\citenamefont {Coolen}(2005)}]{coolen2005mathematical}%
  \BibitemOpen
  \bibfield  {author} {\bibinfo {author} {\bibfnamefont {A.~C.~C.}\ \bibnamefont {Coolen}},\ }\href@noop {} {\emph {\bibinfo {title} {The mathematical theory of minority games: statistical mechanics of interacting agents}}}\ (\bibinfo  {publisher} {Oxford University Press, Oxford UK},\ \bibinfo {year} {2005})\BibitemShut {NoStop}%
\bibitem [{\citenamefont {Berg}\ and\ \citenamefont {Engel}(1998)}]{berg}%
  \BibitemOpen
  \bibfield  {author} {\bibinfo {author} {\bibfnamefont {J.}~\bibnamefont {Berg}}\ and\ \bibinfo {author} {\bibfnamefont {A.}~\bibnamefont {Engel}},\ }\href@noop {} {\bibfield  {journal} {\bibinfo  {journal} {Phys. Rev. Lett.}\ }\textbf {\bibinfo {volume} {81}},\ \bibinfo {pages} {4999} (\bibinfo {year} {1998})}\BibitemShut {NoStop}%
\bibitem [{\citenamefont {Fischer}\ and\ \citenamefont {Hertz}(1993)}]{hertz}%
  \BibitemOpen
  \bibfield  {author} {\bibinfo {author} {\bibfnamefont {K.~H.}\ \bibnamefont {Fischer}}\ and\ \bibinfo {author} {\bibfnamefont {J.~A.}\ \bibnamefont {Hertz}},\ }\href@noop {} {\emph {\bibinfo {title} {Spin Glasses}}}\ (\bibinfo  {publisher} {Cambridge University Press, Cambridge UK},\ \bibinfo {year} {1993})\BibitemShut {NoStop}%
\bibitem [{\citenamefont {May}(1972)}]{may1972will}%
  \BibitemOpen
  \bibfield  {author} {\bibinfo {author} {\bibfnamefont {R.~M.}\ \bibnamefont {May}},\ }\href@noop {} {\bibfield  {journal} {\bibinfo  {journal} {Nature}\ }\textbf {\bibinfo {volume} {238}},\ \bibinfo {pages} {413} (\bibinfo {year} {1972})}\BibitemShut {NoStop}%
\bibitem [{\citenamefont {Allesina}\ and\ \citenamefont {Tang}(2012)}]{allesina2012stability}%
  \BibitemOpen
  \bibfield  {author} {\bibinfo {author} {\bibfnamefont {S.}~\bibnamefont {Allesina}}\ and\ \bibinfo {author} {\bibfnamefont {S.}~\bibnamefont {Tang}},\ }\href@noop {} {\bibfield  {journal} {\bibinfo  {journal} {Nature}\ }\textbf {\bibinfo {volume} {483}},\ \bibinfo {pages} {205} (\bibinfo {year} {2012})}\BibitemShut {NoStop}%
\bibitem [{\citenamefont {Opper}\ and\ \citenamefont {Diederich}(1992)}]{opper1992phase}%
  \BibitemOpen
  \bibfield  {author} {\bibinfo {author} {\bibfnamefont {M.}~\bibnamefont {Opper}}\ and\ \bibinfo {author} {\bibfnamefont {S.}~\bibnamefont {Diederich}},\ }\href@noop {} {\bibfield  {journal} {\bibinfo  {journal} {Physical review letters}\ }\textbf {\bibinfo {volume} {69}},\ \bibinfo {pages} {1616} (\bibinfo {year} {1992})}\BibitemShut {NoStop}%
\bibitem [{\citenamefont {Yoshino}\ \emph {et~al.}(2007)\citenamefont {Yoshino}, \citenamefont {Galla},\ and\ \citenamefont {Tokita}}]{yoshino1}%
  \BibitemOpen
  \bibfield  {author} {\bibinfo {author} {\bibfnamefont {Y.}~\bibnamefont {Yoshino}}, \bibinfo {author} {\bibfnamefont {T.}~\bibnamefont {Galla}}, \ and\ \bibinfo {author} {\bibfnamefont {K.}~\bibnamefont {Tokita}},\ }\href@noop {} {\bibfield  {journal} {\bibinfo  {journal} {Journal of Statistical Mechanics: Theory and Experiment}\ }\textbf {\bibinfo {volume} {2007}},\ \bibinfo {pages} {P09003} (\bibinfo {year} {2007})}\BibitemShut {NoStop}%
\bibitem [{\citenamefont {Yoshino}\ \emph {et~al.}(2008)\citenamefont {Yoshino}, \citenamefont {Galla},\ and\ \citenamefont {Tokita}}]{yoshino2}%
  \BibitemOpen
  \bibfield  {author} {\bibinfo {author} {\bibfnamefont {Y.}~\bibnamefont {Yoshino}}, \bibinfo {author} {\bibfnamefont {T.}~\bibnamefont {Galla}}, \ and\ \bibinfo {author} {\bibfnamefont {K.}~\bibnamefont {Tokita}},\ }\href@noop {} {\bibfield  {journal} {\bibinfo  {journal} {Phys. Rev. E}\ }\textbf {\bibinfo {volume} {78}},\ \bibinfo {pages} {031924} (\bibinfo {year} {2008})}\BibitemShut {NoStop}%
\bibitem [{\citenamefont {Bunin}(2016)}]{bunin2016interaction}%
  \BibitemOpen
  \bibfield  {author} {\bibinfo {author} {\bibfnamefont {G.}~\bibnamefont {Bunin}},\ }\href@noop {} {\bibfield  {journal} {\bibinfo  {journal} {arXiv preprint arXiv:1607.04734}\ } (\bibinfo {year} {2016})}\BibitemShut {NoStop}%
\bibitem [{\citenamefont {Galla}(2018)}]{galla2018dynamically}%
  \BibitemOpen
  \bibfield  {author} {\bibinfo {author} {\bibfnamefont {T.}~\bibnamefont {Galla}},\ }\href@noop {} {\bibfield  {journal} {\bibinfo  {journal} {EPL (Europhysics Letters)}\ }\textbf {\bibinfo {volume} {123}},\ \bibinfo {pages} {48004} (\bibinfo {year} {2018})}\BibitemShut {NoStop}%
\bibitem [{\citenamefont {Biroli}\ \emph {et~al.}(2018)\citenamefont {Biroli}, \citenamefont {Bunin},\ and\ \citenamefont {Cammarota}}]{biroli2018marginally}%
  \BibitemOpen
  \bibfield  {author} {\bibinfo {author} {\bibfnamefont {G.}~\bibnamefont {Biroli}}, \bibinfo {author} {\bibfnamefont {G.}~\bibnamefont {Bunin}}, \ and\ \bibinfo {author} {\bibfnamefont {C.}~\bibnamefont {Cammarota}},\ }\href@noop {} {\bibfield  {journal} {\bibinfo  {journal} {New Journal of Physics}\ }\textbf {\bibinfo {volume} {20}},\ \bibinfo {pages} {083051} (\bibinfo {year} {2018})}\BibitemShut {NoStop}%
\bibitem [{\citenamefont {Altieri}\ \emph {et~al.}(2021)\citenamefont {Altieri}, \citenamefont {Roy}, \citenamefont {Cammarota},\ and\ \citenamefont {Biroli}}]{altieri2021properties}%
  \BibitemOpen
  \bibfield  {author} {\bibinfo {author} {\bibfnamefont {A.}~\bibnamefont {Altieri}}, \bibinfo {author} {\bibfnamefont {F.}~\bibnamefont {Roy}}, \bibinfo {author} {\bibfnamefont {C.}~\bibnamefont {Cammarota}}, \ and\ \bibinfo {author} {\bibfnamefont {G.}~\bibnamefont {Biroli}},\ }\href@noop {} {\bibfield  {journal} {\bibinfo  {journal} {Physical Review Letters}\ }\textbf {\bibinfo {volume} {126}},\ \bibinfo {pages} {258301} (\bibinfo {year} {2021})}\BibitemShut {NoStop}%
\bibitem [{\citenamefont {Roy}\ \emph {et~al.}(2019)\citenamefont {Roy}, \citenamefont {Biroli}, \citenamefont {Bunin},\ and\ \citenamefont {Cammarota}}]{roy2019numerical}%
  \BibitemOpen
  \bibfield  {author} {\bibinfo {author} {\bibfnamefont {F.}~\bibnamefont {Roy}}, \bibinfo {author} {\bibfnamefont {G.}~\bibnamefont {Biroli}}, \bibinfo {author} {\bibfnamefont {G.}~\bibnamefont {Bunin}}, \ and\ \bibinfo {author} {\bibfnamefont {C.}~\bibnamefont {Cammarota}},\ }\href@noop {} {\bibfield  {journal} {\bibinfo  {journal} {Journal of Physics A: Mathematical and Theoretical}\ }\textbf {\bibinfo {volume} {52}},\ \bibinfo {pages} {484001} (\bibinfo {year} {2019})}\BibitemShut {NoStop}%
\bibitem [{\citenamefont {Altieri}\ and\ \citenamefont {Biroli}(2022)}]{Altieri2022}%
  \BibitemOpen
  \bibfield  {author} {\bibinfo {author} {\bibfnamefont {A.}~\bibnamefont {Altieri}}\ and\ \bibinfo {author} {\bibfnamefont {G.}~\bibnamefont {Biroli}},\ }\href@noop {} {\bibfield  {journal} {\bibinfo  {journal} {SciPost Phys.}\ }\textbf {\bibinfo {volume} {12}},\ \bibinfo {pages} {013} (\bibinfo {year} {2022})}\BibitemShut {NoStop}%
\bibitem [{\citenamefont {Poley}\ \emph {et~al.}(2023)\citenamefont {Poley}, \citenamefont {Baron},\ and\ \citenamefont {Galla}}]{Poley}%
  \BibitemOpen
  \bibfield  {author} {\bibinfo {author} {\bibfnamefont {L.}~\bibnamefont {Poley}}, \bibinfo {author} {\bibfnamefont {J.~W.}\ \bibnamefont {Baron}}, \ and\ \bibinfo {author} {\bibfnamefont {T.}~\bibnamefont {Galla}},\ }\href@noop {} {\bibfield  {journal} {\bibinfo  {journal} {Phys. Rev. E}\ }\textbf {\bibinfo {volume} {107}},\ \bibinfo {pages} {024313} (\bibinfo {year} {2023})}\BibitemShut {NoStop}%
\bibitem [{\citenamefont {Hopfield}(1982)}]{hopfield1982neural}%
  \BibitemOpen
  \bibfield  {author} {\bibinfo {author} {\bibfnamefont {J.~J.}\ \bibnamefont {Hopfield}},\ }\href@noop {} {\bibfield  {journal} {\bibinfo  {journal} {Proceedings of the national academy of sciences}\ }\textbf {\bibinfo {volume} {79}},\ \bibinfo {pages} {2554} (\bibinfo {year} {1982})}\BibitemShut {NoStop}%
\bibitem [{\citenamefont {Hebb}(1949)}]{hebb-1949}%
  \BibitemOpen
  \bibfield  {author} {\bibinfo {author} {\bibfnamefont {D.~O.}\ \bibnamefont {Hebb}},\ }\href@noop {} {\emph {\bibinfo {title} {The organization of behavior: {A} neuropsychological theory}}}\ (\bibinfo  {publisher} {Wiley},\ \bibinfo {address} {New York},\ \bibinfo {year} {1949})\BibitemShut {NoStop}%
\bibitem [{\citenamefont {MacArthur}(1955)}]{macarthur1955fluctuations}%
  \BibitemOpen
  \bibfield  {author} {\bibinfo {author} {\bibfnamefont {R.}~\bibnamefont {MacArthur}},\ }\href@noop {} {\bibfield  {journal} {\bibinfo  {journal} {ecology}\ }\textbf {\bibinfo {volume} {36}},\ \bibinfo {pages} {533} (\bibinfo {year} {1955})}\BibitemShut {NoStop}%
\bibitem [{\citenamefont {Martino}\ and\ \citenamefont {Marsili}(2006)}]{demartino2006}%
  \BibitemOpen
  \bibfield  {author} {\bibinfo {author} {\bibfnamefont {A.~D.}\ \bibnamefont {Martino}}\ and\ \bibinfo {author} {\bibfnamefont {M.}~\bibnamefont {Marsili}},\ }\href@noop {} {\bibfield  {journal} {\bibinfo  {journal} {Journal of Physics A: Mathematical and General}\ }\textbf {\bibinfo {volume} {39}},\ \bibinfo {pages} {R465} (\bibinfo {year} {2006})}\BibitemShut {NoStop}%
\bibitem [{\citenamefont {Advani}\ \emph {et~al.}(2018)\citenamefont {Advani}, \citenamefont {Bunin},\ and\ \citenamefont {Mehta}}]{advani2018}%
  \BibitemOpen
  \bibfield  {author} {\bibinfo {author} {\bibfnamefont {M.}~\bibnamefont {Advani}}, \bibinfo {author} {\bibfnamefont {G.}~\bibnamefont {Bunin}}, \ and\ \bibinfo {author} {\bibfnamefont {P.}~\bibnamefont {Mehta}},\ }\href@noop {} {\bibfield  {journal} {\bibinfo  {journal} {Journal of Statistical Mechanics: Theory and Experiment}\ }\textbf {\bibinfo {volume} {2018}},\ \bibinfo {pages} {033406} (\bibinfo {year} {2018})}\BibitemShut {NoStop}%
\bibitem [{\citenamefont {MacArthur}\ and\ \citenamefont {Levins}(1967)}]{MacArthur1967}%
  \BibitemOpen
  \bibfield  {author} {\bibinfo {author} {\bibfnamefont {R.}~\bibnamefont {MacArthur}}\ and\ \bibinfo {author} {\bibfnamefont {R.}~\bibnamefont {Levins}},\ }\href@noop {} {\bibfield  {journal} {\bibinfo  {journal} {Am. Nat.}\ }\textbf {\bibinfo {volume} {101}},\ \bibinfo {pages} {377} (\bibinfo {year} {1967})}\BibitemShut {NoStop}%
\bibitem [{\citenamefont {MacArthur}(1970)}]{MacArthur1970}%
  \BibitemOpen
  \bibfield  {author} {\bibinfo {author} {\bibfnamefont {R.}~\bibnamefont {MacArthur}},\ }\href@noop {} {\bibfield  {journal} {\bibinfo  {journal} {Theor. Popul. Biol.}\ }\textbf {\bibinfo {volume} {1}},\ \bibinfo {pages} {1} (\bibinfo {year} {1970})}\BibitemShut {NoStop}%
\bibitem [{\citenamefont {Chesson}(1990)}]{Cheeson1990}%
  \BibitemOpen
  \bibfield  {author} {\bibinfo {author} {\bibfnamefont {P.}~\bibnamefont {Chesson}},\ }\href@noop {} {\bibfield  {journal} {\bibinfo  {journal} {Theor. Popul. Biol.}\ }\textbf {\bibinfo {volume} {37}},\ \bibinfo {pages} {26} (\bibinfo {year} {1990})}\BibitemShut {NoStop}%
\bibitem [{\citenamefont {Galla}(2005{\natexlab{a}})}]{galla_hebbian}%
  \BibitemOpen
  \bibfield  {author} {\bibinfo {author} {\bibfnamefont {T.}~\bibnamefont {Galla}},\ }\href@noop {} {\bibfield  {journal} {\bibinfo  {journal} {Journal of Statistical Mechanics: Theory and Experiment}\ }\textbf {\bibinfo {volume} {2005}},\ \bibinfo {pages} {P11005} (\bibinfo {year} {2005}{\natexlab{a}})}\BibitemShut {NoStop}%
\bibitem [{\citenamefont {Baron}\ \emph {et~al.}(2023)\citenamefont {Baron}, \citenamefont {Jewell}, \citenamefont {Ryder},\ and\ \citenamefont {Galla}}]{baron2022non}%
  \BibitemOpen
  \bibfield  {author} {\bibinfo {author} {\bibfnamefont {J.~W.}\ \bibnamefont {Baron}}, \bibinfo {author} {\bibfnamefont {T.~J.}\ \bibnamefont {Jewell}}, \bibinfo {author} {\bibfnamefont {C.}~\bibnamefont {Ryder}}, \ and\ \bibinfo {author} {\bibfnamefont {T.}~\bibnamefont {Galla}},\ }\href@noop {} {\bibfield  {journal} {\bibinfo  {journal} {Phys. Rev. Lett.}\ }\textbf {\bibinfo {volume} {130}},\ \bibinfo {pages} {137401} (\bibinfo {year} {2023})}\BibitemShut {NoStop}%
\bibitem [{Sup()}]{Supplement}%
  \BibitemOpen
  \href@noop {} {\enquote {\bibinfo {title} {Supplemental material},}\ }\BibitemShut {NoStop}%
\bibitem [{\citenamefont {Marcus~S}(2022)}]{Marcus2022}%
  \BibitemOpen
  \bibfield  {author} {\bibinfo {author} {\bibfnamefont {B.~G.}\ \bibnamefont {Marcus~S}, \bibfnamefont {Turner~AM}},\ }\href@noop {} {\bibfield  {journal} {\bibinfo  {journal} {PLoS Comput. Biol.}\ }\textbf {\bibinfo {volume} {18(7)}},\ \bibinfo {pages} {e1010274} (\bibinfo {year} {2022})}\BibitemShut {NoStop}%
\bibitem [{\citenamefont {Kneitel}(2008)}]{KNEITEL20081731}%
  \BibitemOpen
  \bibfield  {author} {\bibinfo {author} {\bibfnamefont {J.}~\bibnamefont {Kneitel}},\ }in\ \href@noop {} {\emph {\bibinfo {booktitle} {Encyclopedia of Ecology}}},\ \bibinfo {editor} {edited by\ \bibinfo {editor} {\bibfnamefont {S.~E.}\ \bibnamefont {Jørgensen}}\ and\ \bibinfo {editor} {\bibfnamefont {B.~D.}\ \bibnamefont {Fath}}}\ (\bibinfo  {publisher} {Academic Press},\ \bibinfo {address} {Oxford},\ \bibinfo {year} {2008})\ pp.\ \bibinfo {pages} {1731--1734}\BibitemShut {NoStop}%
\bibitem [{\citenamefont {De~Dominicis}(1978)}]{de1978dynamics}%
  \BibitemOpen
  \bibfield  {author} {\bibinfo {author} {\bibfnamefont {C.}~\bibnamefont {De~Dominicis}},\ }\href@noop {} {\bibfield  {journal} {\bibinfo  {journal} {Physical Review B}\ }\textbf {\bibinfo {volume} {18}},\ \bibinfo {pages} {4913} (\bibinfo {year} {1978})}\BibitemShut {NoStop}%
\bibitem [{\citenamefont {Sidhom}\ and\ \citenamefont {Galla}(2020)}]{sidhom2020ecological}%
  \BibitemOpen
  \bibfield  {author} {\bibinfo {author} {\bibfnamefont {L.}~\bibnamefont {Sidhom}}\ and\ \bibinfo {author} {\bibfnamefont {T.}~\bibnamefont {Galla}},\ }\href@noop {} {\bibfield  {journal} {\bibinfo  {journal} {Physical Review E}\ }\textbf {\bibinfo {volume} {101}},\ \bibinfo {pages} {032101} (\bibinfo {year} {2020})}\BibitemShut {NoStop}%
\bibitem [{\citenamefont {Bunin}(2017)}]{bunin2017ecological}%
  \BibitemOpen
  \bibfield  {author} {\bibinfo {author} {\bibfnamefont {G.}~\bibnamefont {Bunin}},\ }\href@noop {} {\bibfield  {journal} {\bibinfo  {journal} {Physical Review E}\ }\textbf {\bibinfo {volume} {95}},\ \bibinfo {pages} {042414} (\bibinfo {year} {2017})}\BibitemShut {NoStop}%
\bibitem [{\citenamefont {Coolen}(2000)}]{coolen2000statistical}%
  \BibitemOpen
  \bibfield  {author} {\bibinfo {author} {\bibfnamefont {A.}~\bibnamefont {Coolen}},\ }\href@noop {} {\bibfield  {journal} {\bibinfo  {journal} {arXiv preprint cond-mat/0006011}\ } (\bibinfo {year} {2000})}\BibitemShut {NoStop}%
\bibitem [{\citenamefont {Verbeiren}(2003)}]{verbeiren2003dilution}%
  \BibitemOpen
  \bibfield  {author} {\bibinfo {author} {\bibfnamefont {T.}~\bibnamefont {Verbeiren}},\ }\href@noop {} {\bibfield  {journal} {\bibinfo  {journal} {PhD thesis, KU Leuven}\ } (\bibinfo {year} {2003})}\BibitemShut {NoStop}%
\bibitem [{\citenamefont {Eissfeller}\ and\ \citenamefont {Opper}(1992)}]{eissfeller1992new}%
  \BibitemOpen
  \bibfield  {author} {\bibinfo {author} {\bibfnamefont {H.}~\bibnamefont {Eissfeller}}\ and\ \bibinfo {author} {\bibfnamefont {M.}~\bibnamefont {Opper}},\ }\href@noop {} {\bibfield  {journal} {\bibinfo  {journal} {Physical review letters}\ }\textbf {\bibinfo {volume} {68}},\ \bibinfo {pages} {2094} (\bibinfo {year} {1992})}\BibitemShut {NoStop}%
\bibitem [{\citenamefont {Galla}(2005{\natexlab{b}})}]{galla2005dynamics}%
  \BibitemOpen
  \bibfield  {author} {\bibinfo {author} {\bibfnamefont {T.}~\bibnamefont {Galla}},\ }\href@noop {} {\bibfield  {journal} {\bibinfo  {journal} {Journal of Statistical Mechanics: Theory and Experiment}\ }\textbf {\bibinfo {volume} {2005}},\ \bibinfo {pages} {P11005} (\bibinfo {year} {2005}{\natexlab{b}})}\BibitemShut {NoStop}%
\bibitem [{\citenamefont {Sommers}\ \emph {et~al.}(1988)\citenamefont {Sommers}, \citenamefont {Crisanti}, \citenamefont {Sompolinsky},\ and\ \citenamefont {Stein}}]{sommers1988spectrum}%
  \BibitemOpen
  \bibfield  {author} {\bibinfo {author} {\bibfnamefont {H.~J.}\ \bibnamefont {Sommers}}, \bibinfo {author} {\bibfnamefont {A.}~\bibnamefont {Crisanti}}, \bibinfo {author} {\bibfnamefont {H.}~\bibnamefont {Sompolinsky}}, \ and\ \bibinfo {author} {\bibfnamefont {Y.}~\bibnamefont {Stein}},\ }\href@noop {} {\bibfield  {journal} {\bibinfo  {journal} {Physical review letters}\ }\textbf {\bibinfo {volume} {60}},\ \bibinfo {pages} {1895} (\bibinfo {year} {1988})}\BibitemShut {NoStop}%
\bibitem [{\citenamefont {Aceituno}\ \emph {et~al.}(2019)\citenamefont {Aceituno}, \citenamefont {Rogers},\ and\ \citenamefont {Schomerus}}]{aceituno2019universal}%
  \BibitemOpen
  \bibfield  {author} {\bibinfo {author} {\bibfnamefont {P.~V.}\ \bibnamefont {Aceituno}}, \bibinfo {author} {\bibfnamefont {T.}~\bibnamefont {Rogers}}, \ and\ \bibinfo {author} {\bibfnamefont {H.}~\bibnamefont {Schomerus}},\ }\href@noop {} {\bibfield  {journal} {\bibinfo  {journal} {Physical Review E}\ }\textbf {\bibinfo {volume} {100}},\ \bibinfo {pages} {010302} (\bibinfo {year} {2019})}\BibitemShut {NoStop}%
\end{thebibliography}%


\begin{thebibliography}{6}%
\makeatletter
\providecommand \@ifxundefined [1]{%
 \@ifx{#1\undefined}
}%
\providecommand \@ifnum [1]{%
 \ifnum #1\expandafter \@firstoftwo
 \else \expandafter \@secondoftwo
 \fi
}%
\providecommand \@ifx [1]{%
 \ifx #1\expandafter \@firstoftwo
 \else \expandafter \@secondoftwo
 \fi
}%
\providecommand \natexlab [1]{#1}%
\providecommand \enquote  [1]{``#1''}%
\providecommand \bibnamefont  [1]{#1}%
\providecommand \bibfnamefont [1]{#1}%
\providecommand \citenamefont [1]{#1}%
\providecommand \href@noop [0]{\@secondoftwo}%
\providecommand \href [0]{\begingroup \@sanitize@url \@href}%
\providecommand \@href[1]{\@@startlink{#1}\@@href}%
\providecommand \@@href[1]{\endgroup#1\@@endlink}%
\providecommand \@sanitize@url [0]{\catcode `\\12\catcode `\$12\catcode `\&12\catcode `\#12\catcode `\^12\catcode `\_12\catcode `\%12\relax}%
\providecommand \@@startlink[1]{}%
\providecommand \@@endlink[0]{}%
\providecommand \url  [0]{\begingroup\@sanitize@url \@url }%
\providecommand \@url [1]{\endgroup\@href {#1}{\urlprefix }}%
\providecommand \urlprefix  [0]{URL }%
\providecommand \Eprint [0]{\href }%
\providecommand \doibase [0]{http://dx.doi.org/}%
\providecommand \selectlanguage [0]{\@gobble}%
\providecommand \bibinfo  [0]{\@secondoftwo}%
\providecommand \bibfield  [0]{\@secondoftwo}%
\providecommand \translation [1]{[#1]}%
\providecommand \BibitemOpen [0]{}%
\providecommand \bibitemStop [0]{}%
\providecommand \bibitemNoStop [0]{.\EOS\space}%
\providecommand \EOS [0]{\spacefactor3000\relax}%
\providecommand \BibitemShut  [1]{\csname bibitem#1\endcsname}%
\let\auto@bib@innerbib\@empty
\bibitem [{\citenamefont {Opper}\ and\ \citenamefont {Diederich}(1992)}]{opper1992phase}%
  \BibitemOpen
  \bibfield  {author} {\bibinfo {author} {\bibfnamefont {M.}~\bibnamefont {Opper}}\ and\ \bibinfo {author} {\bibfnamefont {S.}~\bibnamefont {Diederich}},\ }\href@noop {} {\bibfield  {journal} {\bibinfo  {journal} {Physical review letters}\ }\textbf {\bibinfo {volume} {69}},\ \bibinfo {pages} {1616} (\bibinfo {year} {1992})}\BibitemShut {NoStop}%
\bibitem [{\citenamefont {Galla}(2018)}]{galla2018dynamically}%
  \BibitemOpen
  \bibfield  {author} {\bibinfo {author} {\bibfnamefont {T.}~\bibnamefont {Galla}},\ }\href@noop {} {\bibfield  {journal} {\bibinfo  {journal} {EPL (Europhysics Letters)}\ }\textbf {\bibinfo {volume} {123}},\ \bibinfo {pages} {48004} (\bibinfo {year} {2018})}\BibitemShut {NoStop}%
\bibitem [{\citenamefont {Coolen}(2001)}]{Coolen_dyn}%
  \BibitemOpen
  \bibfield  {author} {\bibinfo {author} {\bibfnamefont {A.~C.~C.}\ \bibnamefont {Coolen}},\ }in\ \href@noop {} {\emph {\bibinfo {booktitle} {Neuro-Informatics and Neural Modelling}}},\ \bibinfo {series} {Handbook of Biological Physics}, Vol.~\bibinfo {volume} {4},\ \bibinfo {editor} {edited by\ \bibinfo {editor} {\bibfnamefont {F.}~\bibnamefont {Moss}}\ and\ \bibinfo {editor} {\bibfnamefont {S.}~\bibnamefont {Gielen}}}\ (\bibinfo  {publisher} {North-Holland},\ \bibinfo {year} {2001})\ pp.\ \bibinfo {pages} {619--684}\BibitemShut {NoStop}%
\bibitem [{\citenamefont {Verbeiren}(2003)}]{verbeiren2003dilution}%
  \BibitemOpen
  \bibfield  {author} {\bibinfo {author} {\bibfnamefont {T.}~\bibnamefont {Verbeiren}},\ }\href@noop {} {\bibfield  {journal} {\bibinfo  {journal} {PhD thesis, KU Leuven}\ } (\bibinfo {year} {2003})}\BibitemShut {NoStop}%
\bibitem [{\citenamefont {Coolen}(2000)}]{coolen2000statistical}%
  \BibitemOpen
  \bibfield  {author} {\bibinfo {author} {\bibfnamefont {A.}~\bibnamefont {Coolen}},\ }\href@noop {} {\bibfield  {journal} {\bibinfo  {journal} {arXiv preprint cond-mat/0006011}\ } (\bibinfo {year} {2000})}\BibitemShut {NoStop}%
\bibitem [{\citenamefont {Coolen}(2005)}]{coolen2005mathematical}%
  \BibitemOpen
  \bibfield  {author} {\bibinfo {author} {\bibfnamefont {A.~C.~C.}\ \bibnamefont {Coolen}},\ }\href@noop {} {\emph {\bibinfo {title} {The mathematical theory of minority games: statistical mechanics of interacting agents}}}\ (\bibinfo  {publisher} {Oxford University Press, Oxford UK},\ \bibinfo {year} {2005})\BibitemShut {NoStop}%
\end{thebibliography}%

\begin{appendix}   
\section{Details of numerical procedures}\label{app:numerics}
For the numerical integration of the gLVE \eqref{HLV} we use scypi's \texttt{solve\_ivp} function, which uses a RK45 integration scheme. 

To determine the fraction of survivors we count the number of species above a threshold abundance of $10^{-4}$. There are two sources of systematic error associated with this method. The most relevant is the overestimation of the fraction of survivors if the system is not close enough to the equilibrium configuration. This can be addressed by extending the simulation time.

The second source of error comes from the fact there is no `gap' between zero and the lowest non-zero abundance [see the clipped Gaussian Eq.~\eqref{fxd_point}]. This implies that for any value of the threshold, there is a nonzero probability of finding equilibrium abundances below it. A possible solution, making use of the facts that in simulations $N$ is finite and that we know the abundance distribution analytically, is to choose the threshold value so that the expected number of surviving species with an abundance below the threshold is small (e.g. smaller than one). The chosen value of $10^{-4}$ provides good results in the parameter ranges we have explored.

As part of our measurements, it is necessary to detect divergences in the species' abundances. To detect this divergence we have used the failure of the integration method as an indicator. Indeed, as the abundances grow with each iteration, so does the estimated error used to adapt the step size. This causes the solver to lower the time-step until it eventually drops below machine precision, at which point integration is stopped. The agreement of the theoretical and numerical phase boundaries in Fig.~\ref{fig:t_div_1} confirms the validity of this method.

\newpage

\section{Order parameters at the fixed point of the fully connected system}\label{app:c_eq_1}

The parametric solution for the order parameters of the fully connected system ($c=1$) in the fixed point phase can be obtained from the following relations,
\begin{align}
\alpha &= \dfrac{(u + f_0)^2 f_2}{(f_0 + f_2)^2}, \nonumber \\
M &= \dfrac{f_1^2 (f_0 + f_2)}{(f_0 - f_2) (f_0^2 - uf_2)}, \nonumber \\
q &= \dfrac{f_1^2 f_2 (f_0 + f_2)^2}{(f_0 - f_2)^2 (f_0^2 - uf_2)^2}, \nonumber \\
\chi &= \dfrac{ f_0 ( f_0+ f_2)}{ f_0^2- uf_2}.\label{eq:c_eq_1}
\end{align}
The functions $f_n(\Delta)$ on the right provide $\alpha, M, q, \chi$ as implicit functions of $\Delta$.

Keeping Eq.~\eqref{eq:thedefinition} in mind one sees that $M$ and $q$ can only diverge if $f_0=f_2$ or $f_0^2=uf_2$, as indicated in the main text.

\section{Limiting behaviour of the order parameters}\label{app:limit}
\subsection{Limit $\alpha\to 0$}
The weak interaction limit $\alpha\rightarrow 0$ corresponds to $\Delta\rightarrow\infty$. [This can be seen from Eq.~(\ref{eq:M_q_alpha}), keeping in mind that $f_0>0$]. From the definition in Eqs.~\eqref{eq:thedefinition} we have, in this limit, 
\begin{align}
	f_0(\Delta) & = \phi = 1 + \frac{e^{-\frac{\Delta^2}{2}}}{\sqrt{2 \pi }}\left[-\frac{1}{ \Delta}+\mathcal{O}\left(\Delta^{-3}\right)\right] \label{eq:see} \\
	f_1(\Delta) & = \Delta + \frac{e^{-\frac{\Delta^2}{2}}}{\sqrt{2 \pi }}\left[\frac{1}{\Delta^2}+\mathcal{O}\left(\Delta^{-4}\right)\right] \\
	f_2(\Delta) & = 1 + \Delta^2 + \frac{e^{-\frac{\Delta^2}{2}}}{\sqrt{2 \pi }}\left[-\frac{2}{\Delta^3}+\mathcal{O}\left(\Delta^{-5}\right)\right].
\end{align}

Next, we compute the value of $\chi$. Since only $f_0$ and $f_2$ are present in Eq.~\eqref{eq:c_less_1}, and only $f_2$ is divergent, we group in terms proportional to $f_2$ to obtain
\begin{equation}
	0=1-\chi\left[2(1-c)-u\right]-\chi^2(c-1)(1-2u)-\chi^3u(c-1), \nonumber
\end{equation}
which we can check always has $\chi=-1/u$ as its negative solution. Finally, from Eq.~\eqref{eq:M_q_alpha} we obtain $M=1/u$ and $\text{Var}[x]=0$.

As expected, these values are independent of $c$ and $\Gamma$, since in the limit of absent interactions Eq.~\eqref{eq:HLV_1} becomes a set of independent logistic maps. In this case, all species survive with abundance $1/u$, which is what we obtain.

\bigskip

\subsection{Limit $\alpha\to\alpha_c$}
There are two different scenarios for the limit $\alpha\rightarrow\alpha_c$ (where $\alpha_c$ is the location of the phase transition). 
\medskip

(1) If $c=1$, $u<1/2$ the divergence takes place as $uf_2\rightarrow f_0^2$. Using Eq.~\eqref{eq:c_eq_1} we see that $\alpha_c=u$, and
\begin{equation}
	\alpha_c-\alpha = \frac{u-f_2}{(f_0+f_2)^2}(f_0^2-uf_2).
\end{equation}
This implies that both $\chi$ and $M$ diverge as $(\alpha_c-\alpha)^{-1}$, and $q$ diverges as $(\alpha_c-\alpha)^{-2}$.
\medskip

(2) The other type of transition occurs when  $f_0\to f_2$, which implies $\Delta \rightarrow 0$. In this case $\chi$ remains finite and we have near $\Delta=0$,
\begin{align}
 f_0(\Delta) & = \frac{1}{2}+\frac{\Delta}{\sqrt{2 \pi }}-\frac{\Delta^3}{6 \sqrt{2 \pi }}+{\cal O}\left(\Delta^4\right) \nonumber\\
 f_1(\Delta) & = \frac{1}{\sqrt{2 \pi }}+\frac{\Delta}{2}+\frac{\Delta^2}{2 \sqrt{2 \pi }}+{\cal O}\left(\Delta^4\right) \nonumber\\
 f_2(\Delta) & = \frac{1}{2}+\sqrt{\frac{2}{\pi }} \Delta+\frac{\Delta^2}{2}+\frac{\Delta^3}{3 \sqrt{2 \pi }}+{\cal O}\left(\Delta^4\right).
\end{align}

We note from these expansions that $f_0-f_2\propto \Delta$ as $\Delta\to 0$. Using Eq.~\eqref{eq:M_q_alpha} we then find $M\sim\Delta^{-1}$. Similarly, we have $\alpha_c-\alpha \propto \Delta$ [this can be seen from expanding $f_0^2/f_2$ in the third relation in Eq.~(\ref{eq:M_q_alpha})], so we can conclude that $M\sim (\alpha-\alpha_c)^{-1}$. These results are consistent with simulations (see for example Fig. \ref{fig:theory_sim}).

\end{appendix}

\end{document}


\title{\edited{Niche overlap and Hopfield-like interactions in generalised random Lotka--Volterra systems}}

\author{Enrique Rozas Garcia}
\email{enrique.rozas.garcia@physics.gu.se}
\affiliation{Department of Physics, Gothenburg University, 41296 Gothenburg, Sweden}
\affiliation{Instituto de F\'isica Interdisciplinar y Sistemas Complejos, IFISC (CSIC-UIB), Campus Universitat Illes Balears, E-07122 Palma de Mallorca, Spain}
\author{Mark J. Crumpton}
\email{mark.j.crumpton@kcl.ac.uk}
\affiliation{Department of Mathematics, King's College London, London WC2R 2LS, United Kingdom}
\affiliation{Department of Physics and Astronomy, School of Natural Sciences, The University of Manchester, Manchester M13 9PL, UK}

\author{Tobias Galla}
\email{tobias.galla@ifisc.uib-csic.es}
\affiliation{Department of Physics and Astronomy, School of Natural Sciences, The University of Manchester, Manchester M13 9PL, UK}
\affiliation{Instituto de F\'isica Interdisciplinar y Sistemas Complejos, IFISC (CSIC-UIB), Campus Universitat Illes Balears, E-07122 Palma de Mallorca, Spain}
\date{\today}

\counterwithout{equation}{section}

\newpage
\onecolumngrid
\clearpage

\setcounter{page}{1}

\begin{center}
\Large{SUPPLEMENTAL MATERIAL\\ Niche overlap and Hopfield-like interactions in generalised random Lotka--Volterra systems}\\ 
\end{center}
\begin{center}
Enrique Rozas Garcia${}^{1,2}$, Mark J. Crumpton${}^{3,4}$, Tobias Galla${}^{1,4}$\\~\\

 ${}^{1}$ Instituto de F\'isica Interdisciplinar y Sistemas Complejos, IFISC (CSIC-UIB), Campus Universitat Illes Balears, E-07122 Palma de Mallorca, Spain\\
 ${}^{2}$ Department of Physics, Gothenburg University, 41296 Gothenburg, Sweden\\
 ${}^{3}$ Department of Mathematics, King's College London, London WC2R 2LS, United Kingdom\\
 ${}^{4}$ Department of Physics and Astronomy, School of Natural Sciences, The University of Manchester, Manchester M13 9PL, UK
 
\end{center}

\hrulefill
\setcounter{figure}{0}
\setcounter{equation}{0}
\setcounter{section}{0}
\setcounter{page}{1}
\renewcommand{\thesection}{S\arabic{section}} 
\renewcommand{\theequation}{S\arabic{equation}}
\renewcommand{\thefigure}{S\arabic{figure}}
\renewcommand{\thepage}{S\arabic{page}}
\fontsize{12}{14.5}\selectfont
\section{Generation of randomly diluted interactions}
\label{Appendix:Bernoulli}

The dynamics studied in this work, of the form in Eq.~(1), requires the generation of pairs of identically distributed correlated Bernoulli random variables $(c_{ij}, c_{ji})$ as part of the simulations. To do this we have to find their joint probability distribution. Let
\begin{align}
    \nonumber &p(c_{ij}=1,c_{ji}=1) = x ,\qquad
    p(c_{ij}=1,c_{ji}=0) = y , \\
    &p(c_{ij}=0,c_{ji}=1) = w ,\qquad
    p(c_{ij}=0,c_{ji}=0) = z.
\end{align}
We can solve for $x,y,w,z$ using the desired moments in Eq. (2) and the normalization condition, i.e. 
\begin{align}
&\langle c_{ij} \rangle = x + y = c \  ,\qquad
\langle c_{ji} \rangle = x + w = c \  ,\qquad \\
&\nonumber \langle c_{ij}c_{ji} \rangle = x = \Gamma c(1-c) + c^2 \ ,\qquad
x+y+w+z = 1 \ ,
\end{align}
to obtain
\begin{align}
p(c_{ij}=1,c_{ji}=1) &= \Gamma c(1-c) + c^2, \nonumber \\
p(c_{ij}=1,c_{ji}=0) = 	p(c_{ij}=0,c_{ji}=1) &= c - \Gamma c(1-c) - c^2, \nonumber\\
p(c_{ij}=0,c_{ji}=0) &= 1-2c + \Gamma c(1-c) + c^2.
\end{align}

\begin{figure}[h]
	\centering
	\includegraphics[width=0.6\linewidth]{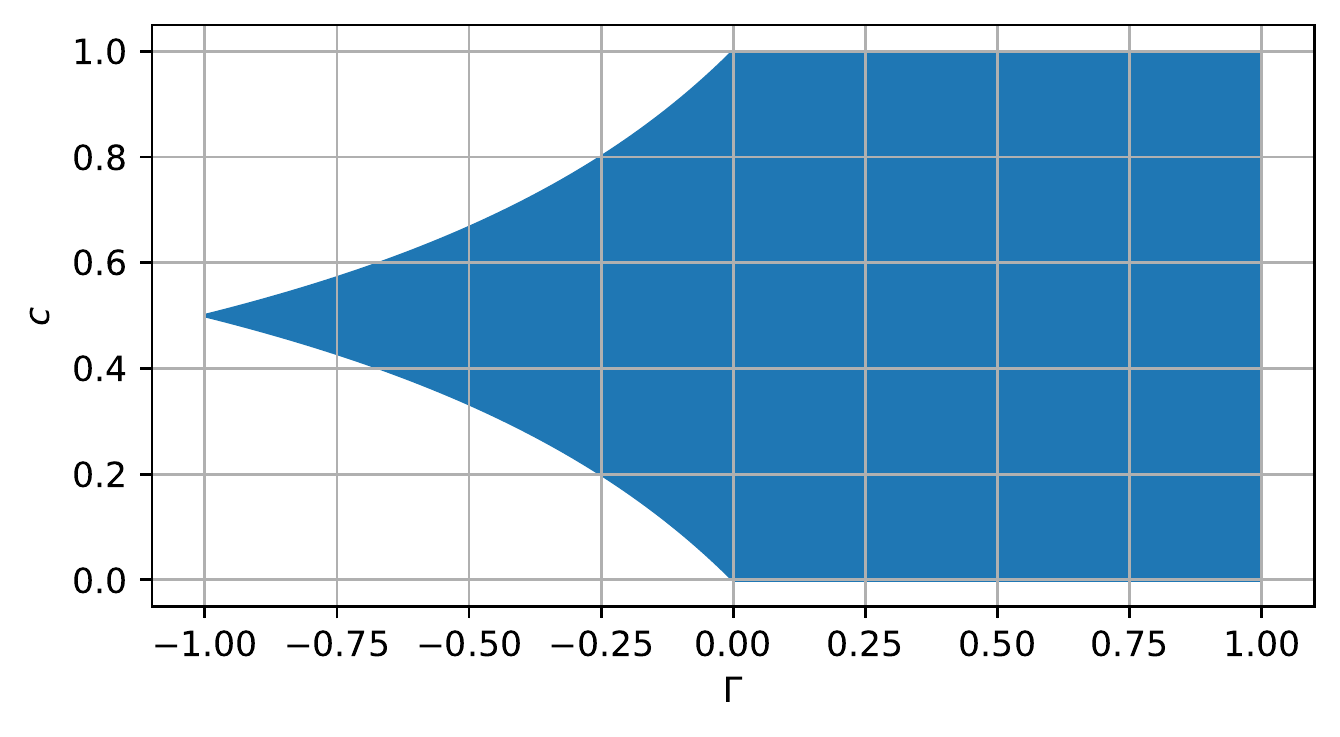}
	\caption{Shaded area indicates the possible pairs of $c$ and $\Gamma$.}
	\label{fig:possiblecorr}
\end{figure}

Since $c\in[0,1]$ and all probabilities need to be non-negative, we find that the values that $\Gamma$ can take are restricted, as shown in Fig. \ref{fig:possiblecorr}. Intuitively, since our variables are identically distributed and can only take two possible values, negative correlations imply exclusion. For example, it is impossible for both variables to have mean $1$ if they are perfectly negatively correlated, in fact, in that case one will have mean $c$ and the other $1-c$, making $c=1/2$ the only possible choice (see Fig. \ref{fig:possiblecorr}).

\section{Details of generating functional analysis}

\subsection{The generating functional}

To study Lotka-Volterra equations of the form in Eq.~(1), we will follow \cite{opper1992phase, galla2018dynamically} and use the generating functional
\begin{align}
    \mathcal{Z}[\bm{\psi}(t)] = \int&\mathcal{D}[\bm{x}(t)]\, p[\bm{x}(0)]e^{i\sum_i\int dt \psi_i(t)x_i(t)} \\
    \nonumber &\prod_{i=1}^N \delta\left[\frac{d}{dt}\ln x_i-1+ux_i-\sum_{i\neq j} c_{ij}J_{ij}x_j +\sigma\zeta_i(t) + h_i(t)\right],
\end{align}
where ${\bm{\psi}}$ is a source field. We have introduced both a Gaussian white noise $\zeta_i(t)$ and a perturbation field $h_i(t)$. These auxiliary fields will allow us to calculate magnitudes of interest later on, but they do not play any role in the upcoming calculation and can be ignored for the most part.

Writing the $\delta$-functions in their integral representation (and absorbing any resulting factors in the integration measure) we find
\begin{equation}
\mathcal{Z}[\bm{\psi}(t) ] = 
\int\mathcal{D}[\bm{x}(t), \bm{\hat{x}}(t)]p[\bm{x}(0)]\,
e^{i\sum_i\int dt\, \psi_i(t)x_i(t)}
e^{i\sum_i\int dt\,\hat{x}_i\left[y_i-1+ux_i-\sum_{j\neq i} c_{ij}J_{ij}x_j + \sigma\zeta_i(t)+h_i(t)\right]} \label{ED1}
\end{equation}
where we use the short-hand $y_i =\frac{d}{dt}\ln x_i$.

As an aside we note that one can alternatively proceed by introducing integration variables $f_i(t)$ and $\hat f_i(t)$, and then start from a generating functional of the form
\begin{eqnarray}
\mathcal{Z}[\bm{\psi}(t) ] &=& 
\int \mathcal{D}[\bm{f}(t), \bm{\hat{f}}(t)]\mathcal{D}[\bm{x}(t), \bm{\hat{x}}(t)]p[\bm{x}(0)]\,
e^{i\sum_i\int dt\, \psi_i(t)x_i(t)} \nonumber \\
&& \times
e^{i\sum_i\int dt\,\hat{x}_i\left[\dot x_i-x_i(t)f_i(t)\right]} e^{i\sum_i\int dt\,\hat{f}_i(t)\left[f_i(t)-1+ux_i-\sum_{j} c_{ij}J_{ij}x_j+\sigma\zeta_i(t)+h_i(t)\right]}.
\end{eqnarray}
One can then follow lines very similar to what is described below, and, after undoing the transformation, arrives at the same final result for the dynamic mean field process for the model. In the text below we use the generating functional in Eq.~\eqref{ED1} as a starting point.
\medskip

Our objective is thus to manipulate the generating functional \eqref{ED1} until we can identify a simplified effective dynamics in terms of a single mean-field population $x(t)$. 

\medskip

To do this we calculate the disorder average of $\mathcal{Z}$. This consists of an average over both random variables $\{c_{ij}\}$ and $\left\{\xi_{i}^\mu\right\}$:
\begin{align}
	\overline{\mathcal{Z}}[\bm{\psi}(t)] &= 
	\int\mathcal{D}[\bm{x}(t), \bm{\hat{x}}(t)]p[\bm{x}(0)]\,
	e^{i\sum_i\int dt\, \psi_i(t)x_i(t)}
	e^{i\sum_i\int dt\,\hat{x}_i\left[y_i-1+ux_i+\sigma\zeta_i(t)+h_i(t)\right]} \nonumber \\
    &\qquad \times
	\overline{e^{-i\sum_i\int dt\,\hat{x}_i\left[\sum_{j\neq i} c_{ij}J_{ij}x_j\right]}},\label{tdf}
\end{align}
where the overbar denotes the disorder average.

\medskip

Once the disorder-averaged generating functional is known, we can calculate magnitudes of interest by taking derivatives. For example, the mean abundance, correlation function, and response function can be written as
	\begin{align}
	M(t) & = \frac{1}{N}\sum_i \overline{\langle x_i(t)\rangle} =  \frac{-i}{N} \sum_i \frac{\delta\overline{\mathcal{Z}}}{\delta \psi_i(t)} \biggr\lvert_{\bm{\psi}(t)  = 0} \label{start}\\
	C(t, t') &= \frac{1}{N}\sum_i \overline{\langle x_i(t) x_i(t')\rangle} = \frac{-1}{N}\sum_i \frac{\delta^2\overline{\mathcal{Z}}}{\delta \psi_i(t)\delta \psi_i(t')}\biggr\lvert_{\bm{\psi}(t)  = 0} \\
	G(t, t') &= \frac{1}{N}\sum_i \frac{\delta\overline{\langle x_i(t)\rangle}}{\delta h_i(t')} =  \frac{-i}{N}\sum_i \frac{\delta^2\overline{\mathcal{Z}}}{\delta \psi_i(t)\delta h_i(t')} \biggr\lvert_{\bm{\psi}(t)  = 0}. \label{end}
	\end{align}
 
\subsection{Introduction of the order parameters}
Before we explicitly carry out the disorder average it is convenient to introduce the following order parameters
\begin{align}
    a(t) &= \frac{1}{N}\sum_{i} x_i(t) \nonumber \\
    k(t) &= i\frac{1}{N}\sum_{i} \hat{x}_i(t)  \nonumber\\
    Q(t,t') &=  \frac{1}{N}\sum_{i}x_i(t)x_i(t')  \nonumber\\
    K(t,t')) &=  \frac{1}{N}\sum_{i}x_i(t)\hat{x}_i(t') \nonumber \\
    L(t,t') &=  \frac{1}{N}\sum_{i}\hat{x}_i(t)\hat{x}_i(t').
\end{align}

To introduce the order parameters into the generating functional \eqref{tdf} we use $\delta$-functions in their exponential representation, and insert the following expressions (all equal to one),
\begin{align}
1 &= \int \mathcal{D}[\hat{a}(t), a(t)] \, e^{iN\int dt\, \hat{a}(t)\left[a(t)-\frac{1}{N}\sum_i x_i(t)\right],} \nonumber \\
1 &= \int \mathcal{D}[\hat{k}(t), k(t)] \, e^{iN\int dt\, \hat{k}(t)\left[k(t)-\frac{i}{N}\sum_i \hat{x}_i(t)\right]}, \nonumber \\
1 &= \int \mathcal{D}[\hat{Q}(t, t'), Q(t, t')] \, e^{iN\int dt\, dt'\, \hat{Q}(t, t')\left[Q(t, t')-\frac{1}{N}\sum_i x_i(t)x_i(t')\right]}, \nonumber \\
1 &= \int \mathcal{D}[\hat{K}(t, t'), K(t, t')] \, e^{iN\int dt\, dt'\, \hat{K}(t, t')\left[K(t, t')-\frac{1}{N}\sum_i x_i(t)\hat{x}_i(t')\right]},  \nonumber\\
1 &= \int \mathcal{D}[\hat{L}(t, t'), L(t, t')] \, e^{iN\int dt\, dt'\, \hat{L}(t, t')\left[L(t, t')-\frac{1}{N}\sum_i \hat{x}_i(t)\hat{x}_i(t')\right]}.
\end{align}

Relevant factors of $2\pi$ have here been absorbed into the measure. By introducing these expressions into \eqref{tdf} we obtain
\begin{align}
\overline{\mathcal{Z}}[\bm{\psi}(t)] = & \int \mathcal{D}[a, \hat{a}, k, \hat{k}, Q, \hat{Q}, K, \hat{K}, L, \hat{L}] \, e^{N\Psi}  	\nonumber \\
&\times \int\mathcal{D}[\bm{x}(t), \bm{\hat{x}}(t)]p[\bm{x}(0)]\,
e^{i\sum_i\int dt\, \psi_i(t)x_i(t)}
e^{i\sum_i\int dt\,\hat{x}_i\left[y_i-1+ux_i+\sigma\zeta_i(t)+h_i(t)\right]} \nonumber \\
& \times e^{-i\sum_i \int dt\, [\hat{a}(t) x_i(t) + i\hat{k}(t)\hat{x}_i(t)]}
e^{-i\sum_i \int dt\,dt'\, [\hat{Q}(t,t') x_i(t)x_i(t') + \hat{K}(t, t')x_i(t)\hat{x}_i(t') + \hat{L}(t, t')\hat{x}_i(t)\hat{x}_i(t')]} \nonumber \\
&\times \overline{e^{-i\sum_i\int dt\,\hat{x}_i\left[\sum_{j\neq i} c_{ij}J_{ij}x_j\right]}},
\end{align}
where 
\begin{align}
    \Psi[a, \hat{a}, k, \hat{k}, Q, \hat{Q}, K, \hat{K}, L, \hat{L}] =& i\int dt\, \left[ \hat{a}(t)a(t) + \hat{k}(t)k(t) \right]\label{Psi} \\
    \nonumber &+ i \int dt\, dt'\, \left[ \hat{Q}(t,t')Q(t,t') + \hat{K}(t,t')K(t,t') + \hat{L}(t,t')L(t,t')\right].
\end{align}

For later purposes we define
\begin{align}
    \Omega[\hat{a}, \hat{k}, \hat{Q}, \hat{K}, \hat{L}]  & = \frac{1}{N}\sum_i \log\bigg[ \int\mathcal{D}[x_i(t), \hat{x}_i(t)]p[x_i(0)]\, 
    \exp\left(i\int dt\, \psi_i(t)x_i(t)\right) \nonumber \\
    & \times \exp\left(i\int dt\,\hat{x}_i\left[y_i-1+ux_i+\sigma\zeta_i(t)+h_i(t)\right]\right) \nonumber \\
    & \times \exp\left(-i \int dt\, [\hat{a}(t) x_i(t) + i\hat{k}(t)\hat{x}_i(t)]\right) \nonumber \\
    & \times \exp\left(-i \int dt\,dt'\, [\hat{Q}(t,t') x_i(t)x_i(t') + \hat{K}(t, t')x_i(t)\hat{x}_i(t') + \hat{L}(t, t')\hat{x}_i(t)\hat{x}_i(t')]\right)\Biggr].  \nonumber \\  \label{Omega}
\end{align}

\subsection{Disorder average}
We follow mainly \cite{Coolen_dyn, verbeiren2003dilution}. The only term in Eq.~\eqref{ED1} containing the disorder variables $\{c_{ij}\}$ and $\left\{\xi^\mu_i\right\}$ is
\begin{equation}
\exp\left[-i\sum_{i>j}\int dt\, (\hat{x}_i c_{ij}J_{ij }x_j+\hat{x}_j c_{ji}J_{ji} x_i) \right]. \label{da2}
\end{equation}
To simplify the calculations we will only keeps the terms that are leading order in $N$, since we eventually intend to take the thermodynamic limit $N\rightarrow\infty$. To estimate the order of each term we use their variance, thus 
\begin{equation}
c_{ij}\sim\sqrt{\langle c_{ij}^2\rangle}=\sqrt{c}\sim\mathcal{O}(N^{0}),
\end{equation}
and 
\begin{equation}
    \langle J_{ij}^2\rangle = \left\langle \left(\frac{1}{c N}\sum_{\mu=1}^{\alpha c N}\xi_i^\mu\xi_j^\mu\right)^2 \right\rangle 
    = \frac{1}{(c N)^2}\sum_{\mu,\mu'=1}^{\alpha c N} \langle \xi_i^\mu\xi_j^\mu\xi_i^{\mu'}\xi_j^{\mu'} \rangle = \frac{1}{(c N)^2}\sum_{\mu,\mu'=1}^{\alpha cN} \delta_{\mu,\mu'} = \frac{\alpha}{c N} \ ,
\end{equation}
implying that
\begin{equation}
    J_{ij}\sim\mathcal{O}(N^{-1/2}) \ .
\end{equation}

We first perform the average of \eqref{da2} over the $\{c_{ij}\}$. We begin by Taylor expanding the exponential
\begin{equation}
\prod_{i< j} e^{-i J_{ij}(b_{ij}c_{ij}+b_{ji}c_{ji})} 
= 
\prod_{i< j}\left[1 - iJ_{ij} (b_{ij}c_{ij}+b_{ji}c_{ji}) - \frac{J_{ij}^2}{2} (b_{ij}c_{ij}+b_{ji}c_{ji})^2 + \mathcal{O}\left(N^{-3/2}\right) \right],
\end{equation}
where we have used $J_{ij}=J_{ji}$, which follows directly from Eq. (3), and we have introduced the abbreviation ${b_{ij}=\int dt\, \hat{x}_i x_j}$. The different factors in the product are independent (since $i<j$) so we can average over $\{c_{ij}\}$ directly to obtain
\begin{align}
&\prod_{i< j}\left[1 - icJ_{ij} (b_{ij}+b_{ji}) - \frac{J_{ij}^2}{2} (b_{ij}^2c+b_{ji}^2c+2b_{ij}b_{ji}[\Gamma c(1-c)+c^2]) \right] \nonumber \\
&=\prod_{i< j}\Bigg[1 
- icJ_{ij} (b_{ij}+b_{ji}) 
+ \frac{1}{2}\left[icJ_{ij} (b_{ij}+b_{ji})\right]^2 
- \frac{1}{2}\left[icJ_{ij} (b_{ij}+b_{ji})\right]^2 \nonumber \\ 
&\hspace{7cm}- \frac{J_{ij}^2}{2}\left(b_{ij}^2c+b_{ji}^2c+2b_{ij}b_{ji}[\Gamma c(1-c)+c^2]\right) \Bigg] \nonumber \\
&=\prod_{i< j}\bigg[1 
- icJ_{ij} (b_{ij}+b_{ji}) 
+ \frac{1}{2}\left[icJ_{ij} (b_{ij}+b_{ji})\right]^2 
- \frac{J_{ij}^2}{2} \left[c(1-c)(b_{ij}+b_{ji})^2\right. \nonumber \\
&\hspace{7cm}\left.+2b_{ij}b_{ji}[\Gamma c(1-c)-c(1-c)]\right] \bigg] \nonumber \\
&= \prod_{i< j}\exp\left[-icJ_{ij} (b_{ij}+b_{ji})- \frac{J_{ij}^2}{2} \left[v^2(b_{ij}+b_{ji})^2+2b_{ij}b_{ji}(w^2-v^2)\right]+\dots\right], \label{da3}
\end{align}
where we have re-exponentiated the expansion with the understanding that the result is only valid in the thermodynamic limit, and used the notation $v^2=\text{Var}[c_{ij}]=c(1-c)$ and $w^2=\text{Corr}[c_{ij},c_{ij}]=\Gamma c(1-c)$.

Now we calculate the average with respect to the variables $\left\{\xi^\mu_i\right\}$. To do this we rewrite the exponentials in Eq.~\eqref{da3} as
\begin{align}
&\exp\left[-ic\sum_{i\neq j}J_{ij} b_{ij} \right] 
\exp\left[-\frac{\alpha}{2cN}\sum_{i\neq j} \left[v^2b_{ij}^2+b_{ij}b_{ji}w^2\right]\right]. \label{xi_avg}
\end{align}
were we have set $J_{ij}^2=\langle J_{ij}^2\rangle$ in the second term (keeping in mind that we are only interested in leading-order terms in the thermodynamic limit). 

Again, retaining only leading-order terms, we can express the different combinations of $b_{ij}$ using the order parameters and find 
\begin{align}
    \sum_{i\neq j} b_{ij}^2 & =  \int dt \, dt'\, \sum_{i\neq j} \hat{x}_i(t)\hat{x}_i(t')x_j(t) x_j(t') \nonumber \\ 
    &= \int dt \, dt'\, \left[\sum_{i, j} \hat{x}_i(t)\hat{x}_i(t')x_j(t) x_j(t') - \sum_i \hat{x}_i(t)\hat{x}_i(t')x_i(t) x_i(t') \right] \nonumber \\
    &= \int dt \, dt'\, \left[ N^2 L(t,t')Q(t,t') + \mathcal{O}(N) \right],
\end{align}
and
\begin{align}
\sum_{i\neq j} b_{ij}b_{ji} & = \int dt \, dt'\, \sum_{i\neq j} \hat{x}_i(t) x_j(t) \hat{x}_j(t') x_i(t') \nonumber \\
&= \int dt \, dt'\, \left[ \sum_{i, j} \hat{x}_i(t) x_j(t) \hat{x}_j(t') x_i(t') - \sum_{i} \hat{x}_i(t) x_i(t) \hat{x}_i(t') x_i(t') \right]\nonumber \\
& = \int dt \, dt'\, \left[ N^2 K(t,t')K(t',t) + \mathcal{O}(N) \right].
\end{align}
To leading order in $N$, the second term in Eq.~\eqref{xi_avg} is therefore
\begin{equation}
\exp\left[-\frac{N\alpha}{2c}\int dt\, dt'\, \left(v^2L(t,t')Q(t,t') + w^2 K(t,t')K(t',t)\right)\right].
\end{equation}

The average of the first term in Eq.~\eqref{xi_avg} can be written as 
\begingroup
\allowdisplaybreaks
\begin{align}
    \left\langle\exp\left[-ic\sum_{i\neq j}J_{ij} b_{ij} \right]\right\rangle_\xi & 
    = \left\langle \exp\left[i\frac{1}{N}\sum_{i\neq j}\sum_{\mu=1}^{\alpha c N}\xi_i^\mu\xi_j^\mu b_{ij} \right] \right\rangle_{\xi} \nonumber\\
    &= \left\langle \exp\left[i\frac{1}{N}\sum_{i,j}\sum_{\mu=1}^{\alpha c N}\xi_i^\mu\xi_j^\mu b_{ij} - i\frac{1}{N}\sum_{i}\sum_{\mu=1}^{\alpha c N}\xi_i^\mu\xi_i^\mu b_{ii}  \right] \right\rangle_{\xi} \nonumber \\
    &= \left\langle \exp\left[i\frac{1}{N}\sum_{i,j}\sum_{\mu=1}^{\alpha c N}\xi_i^\mu\xi_j^\mu b_{ij} - i\alpha c\sum_{i} b_{ii}  \right] \right\rangle_{\xi} 
    \nonumber \\
    &= e^{-i\alpha c\int dt\, \sum_i\hat{x}_i(t)x_i(t)}\left\langle \exp\left[i\frac{1}{N}\sum_{i,j}\sum_{\mu=1}^{\alpha cN}\xi_i^\mu\xi_j^\mu b_{ij} \right] \right\rangle_{\xi} \nonumber \\
    &= e^{-i\alpha cN \int dt\, K(t,t)}\left\langle \exp\left[i\frac{1}{N}\int dt\, \sum_{i,j} \xi_i \xi_j \hat{x}_i(t)x_j(t) \right] \right\rangle_{\xi}^{\alpha cN} .
\end{align}
\endgroup
The sum over $\mu$ in the penultimate line produces $\alpha c N$ identical factors, as indicated in the last line.

At this point, we have   
\begin{align}
    \overline{e^{-i\sum_i\int dt\,\hat{x}_i\left[\sum_{j\neq i} c_{ij}J_{ij}x_j\right]}} =& \frac{1}{N}\log\Bigg[e^{-\frac{N\alpha}{2c}\int dt\, dt'\, \left(v^2L(t,t')Q(t,t') + w^2 K(t,t')K(t',t)\right)}  \nonumber \\
    & \times  
    e^{-i\alpha cN \int dt\, K(t,t)} 
    \left\langle \exp\left[i\frac{1}{N}\int dt\, \sum_{i,j} \xi_i \xi_j \hat{x}_i(t)x_j(t) \right] \right\rangle_\xi^{\alpha cN} \Bigg] \nonumber \\
    =& -\frac{\alpha v^2}{2c}\int dt\, dt'\, \left[L(t,t')Q(t,t') + \Gamma K(t,t')K(t',t)\right]  \label{b3plus} \\
    & - i\alpha c \int dt\, K(t,t) 
    + \alpha c\log\left\langle \exp\left[i\frac{1}{N}\int dt\, \sum_{i,j} \xi_i \xi_j \hat{x}_i(t)x_j(t) \right] \right\rangle_\xi. \nonumber
\end{align}

We now deal with the average over the remaining $\xi_i$. To do this we write
\begin{align}
    \bigg\langle e&^{\frac{i}{N}\int dt\sum_{i,j} \xi_i \xi_j \hat{x}_i x_j } \bigg\rangle_\xi  = \left\langle e^{-i\int dt\, \left(\sum_{i} \frac{\xi_i}{\sqrt{N}} \hat{x}_i\right) \left(-\sum_{j} \frac{\xi_j}{\sqrt{N}} x_j\right)} \right\rangle_\xi \nonumber \\
    & = \int \mathcal{D}[z(t), w(t)] e^{-i\int dt\, z(t)w(t)} \left\langle \prod_t\left\{\delta\left[z(t)-\sum_i\frac{\xi_i}{\sqrt{N}} \hat{x}_i\right]\delta\left[w(t)+\sum_i\frac{\xi_i}{\sqrt{N}} x_i\right]\right\} \right\rangle_\xi \nonumber \\
    & = \int \mathcal{D}[z, w, \hat{z}, \hat{w}] e^{i\int dt\, \left[\hat{z}(t)z(t) + \hat{w}(t)w - z(t)w(t)\right]} \left\langle
    e^{-i\sum_i\frac{\xi_i}{\sqrt{N}}\int dt\, \left[\hat{z}(t)\hat{x}_i(t)-\hat{w}(t)x_i(t)\right] }\right\rangle_\xi \nonumber \\
    & = \int \mathcal{D}[z, w, \hat{z}, \hat{w}] e^{i\int dt\, \left[\hat{z}(t)z(t) + \hat{w}(t)w - z(t)w(t)\right] + \sum_i \log\left\{\cos\left(\frac{1}{\sqrt{N}}\int dt\, \left[\hat{z}(t)\hat{x}_i(t)-\hat{w}(t)x_i(t)\right]\right)\right\}} \nonumber \\
    & = \int \mathcal{D}[z, w, \hat{z}, \hat{w}] e^{i\int dt\, \left[\hat{z}(t)z(t) + \hat{w}(t)w - z(t)w(t)\right] - \frac{1}{2N}\sum_i \left(\int dt\, \left[\hat{z}(t)\hat{x}_i(t)-\hat{w}(t)x_i(t)\right]\right)^2 + {\mathcal{O}(N^{-2})}} \nonumber \\
    & = \int \mathcal{D}[z, w, \hat{z}, \hat{w}] e^{i\int dt\, \left[\hat{z}(t)z(t) + \hat{w}(t)w - z(t)w(t)\right] - \frac{1}{2} \int dt dt'\, \left[
    \hat{z}(t)\hat{z}(t')L(t,t')+\hat{w}(t)\hat{w}(t')Q(t,t')-2\hat{z}(t)\hat{w}(t')K(t',t)
    \right]}, \label{b22}
\end{align}
where in the last steps we have used the Taylor expansion $\log[\cos(x)]=-x^2/2+\mathcal{O}(x^4)$, and expanded the resulting square to introduce the order parameters. This expression can be further simplified by carrying out the integration in the $z$, $w$ variables as follows
\begin{align}
    \int \frac{dx\, dy}{2\pi} e^{i(x\hat{x}+\hat{y}y-xy)} f(\hat{x},\hat{y}) =& \int dx\, e^{i\hat{x}x}f(\hat{x},\hat{y})\int \frac{dy}{2\pi}e^{iy(\hat{y}-x)} \nonumber\\
    =& \int dx\, e^{i\hat{x}x}f(\hat{x},\hat{y})\delta(\hat{y}-x) = f(\hat{x},\hat{y})e^{i\hat{x}\hat{y}}. 
    \label{b23}
\end{align}

Applying the identity in Eq.~\eqref{b23} to Eq.~\eqref{b22} we obtain
\begin{equation}
	\int\mathcal{D}[u, v]\exp\left\{i\int dt\, u(t)v(t)-\frac{1}{2}\int dt\, dt'\, \left[
	v(t)v(t')L(t,t') + u(t)u(t')Q(t,t') - 2v(t)u(t')K(t',t)\right] \right\}.
\end{equation}

With this we are done with the average, and can substitute into Eq. \eqref{b3plus} to obtain the disorder-averaged generating functional in the form
\begin{equation}
    \overline{\mathcal{Z}} = \int \mathcal{D}[a, \hat{a}, k, \hat{k}, Q, \hat{Q}, K, \hat{K}, L, \hat{L}] \, e^{N(\Psi + \Phi + \Omega)} \label{sp_form},
\end{equation}	
to leading order in $N$ in the exponent, where
\begin{align}
    \Phi[Q, K, L] =& - i\alpha c \int dt\, K(t,t) - \frac{\alpha v^2}{2c}\int dt\, dt'\, \left[L(t,t')Q(t,t') + \Gamma K(t,t')K(t',t)\right] \nonumber \\
    & + \alpha c\log\Biggr(\int\mathcal{D}[u, v]\exp\bigg\{-\frac{1}{2}\int dt\, dt'\, \bigg[
    v(t)v(t')L(t,t') + u(t)u(t')Q(t,t') - \nonumber \\
    & \hspace{2cm} 2v(t)u(t')K(t',t)\bigg] \bigg\} 
    \times\exp\left\{i\int dt\, u(t)v(t)\right\}\Biggr),
\end{align}
with $\Psi$ and $\Omega$ defined in Eqs.~(\ref{Psi}) and (\ref{Omega}), respectively.

\subsection{Saddle-point integration}

We now proceed with the saddle-point integration in Eq.~(\ref{sp_form}) in the limit $N\to\infty$. We have
\begin{align}
0 & = \frac{\delta}{\delta \hat{a}(t)}\left[\Psi + \Phi + \Omega \right]  = ia(t) \nonumber \\
&+ \frac{1}{N}\sum_i \frac{\int\mathcal{D}[x_i, \hat{x}_i]p_i(-ix_i) 
	e^{i\sum_i\int dt\, \big[\psi_i x_i+\hat{x}_i\left[y_i-1+ux_i+\sigma\zeta_i +h_i \right]-\hat{a}  x_i  - i\hat{k} \hat{x}_i \big]-i\sum_i \int dt\,dt'\, [\hat{Q} x_i x_i  + \hat{K}x_i \hat{x}_i  + \hat{L}\hat{x}_i \hat{x}_i ]}}{
	\int\mathcal{D}[x_i, \hat{x}_i]p_i\, 
	e^{i\sum_i\int dt\, \big[\psi_i x_i+\hat{x}_i\left[y_i-1+ux_i+\sigma\zeta_i +h_i \right]-\hat{a}  x_i  - i\hat{k} \hat{x}_i \big]-i\sum_i \int dt\,dt'\, [\hat{Q} x_i x_i  + \hat{K}x_i \hat{x}_i  + \hat{L}\hat{x}_i \hat{x}_i ]}}, \\ \nonumber
\end{align}
where the time dependencies have been removed for clarity, and $p_i\equiv p_i[x_i(0)]$. Since the last term has the form of an average, we introduce the notation $\langle\cdot\rangle_*$ for it. Notice that the values of the order parameters to be used in $\langle\cdot\rangle_*$ will be the particular ones we obtain by solving all saddle-point equations simultaneously.

Since $\Phi$ does not contain any of the conjugate variables, and the form of $\Psi$ is so simple, the remaining conjugate variables are easily calculated in the same way. In the thermodynamic limit we can write
\begin{align}
    \frac{\delta}{\delta \hat{a}(t)}\left[\Psi + \Phi + \Omega \right] =  0 & \implies a(t) = \lim_{N\rightarrow\infty}\frac{1}{N}\sum_i \langle x_i(t)\rangle_* \label{start_op} ,\\
    \frac{\delta}{\delta \hat{k}(t)}\left[\Psi + \Phi + \Omega \right] =  0 & \implies k(t) = \lim_{N\rightarrow\infty}\frac{i}{N}\sum_i \langle \hat{x}_i(t)\rangle_*, \\
    \frac{\delta}{\delta \hat{Q}(t,t')}\left[\Psi + \Phi + \Omega \right] =  0 & \implies Q(t, t') = \lim_{N\rightarrow\infty}\frac{1}{N}\sum_i \langle x_i(t)x_i(t')\rangle_*, \\
    \frac{\delta}{\delta \hat{K}(t, t')}\left[\Psi + \Phi + \Omega \right] =  0 & \implies K(t, t') = \lim_{N\rightarrow\infty}\frac{1}{N}\sum_i \langle x_i(t)\hat{x}_i(t')\rangle_*,\\
    \frac{\delta}{\delta \hat{L}(t, t')}\left[\Psi + \Phi + \Omega \right] =  0 & \implies L(t, t') = \lim_{N\rightarrow\infty}\frac{1}{N}\sum_i \langle\hat{x}_i(t)\hat{x}_i(t')\rangle_* \ . \label{end_op}
\end{align}

We can relate Eqs. (\ref{start_op}-\ref{end_op}) with the magnitudes defined in Eqs. (\ref{start}-\ref{end}) by using saddle-point integration again. For example:
\begin{align}
    \lim_{N\rightarrow\infty} m(t) & = \lim_{N\rightarrow\infty} \frac{-i}{N} \sum_i \frac{\delta\overline{\mathcal{Z}}}{\delta \psi_i(t)} \biggr\lvert_{\bm{\psi}(t)  = 0} \nonumber \\ &=\lim_{N\rightarrow\infty} -i \sum_i 
    \frac{\int \mathcal{D}[a, \hat{a}, k, \hat{k}, Q, \hat{Q}, K, \hat{K}, L, \hat{L}] \, e^{N(\Psi + \Phi + \Omega))} \frac{\delta \Omega}{\delta \psi_i(t)}}{\int \mathcal{D}[a, \hat{a}, k, \hat{k}, Q, \hat{Q}, K, \hat{K}, L, \hat{L}] \, e^{N(\Psi + \Phi + \Omega))}}
    \biggr\lvert_{\bm{\psi}(t)  = 0} \nonumber \\
    & = \lim_{N\rightarrow\infty}-i \sum_i \frac{\delta \Omega}{\delta \psi_i(t)} \biggr \lvert_{\text{saddle-point}, \bm{\psi}(t)  = 0} = \lim_{N\rightarrow\infty}\frac{1}{N}\sum_i\langle x_i(t)\rangle_* \lvert_{\bm{\psi}(t)  = 0} \ ,
\end{align}
where in the second step we have divided by $\mathcal{Z}[0]$ to take into account the overall constant in the integration measure. The same calculations can be used on the other parameters to obtain useful relations
\begin{align}
m(t) &= \lim_{N\rightarrow\infty} \frac{-i}{N} \sum_i \frac{\delta\overline{\mathcal{Z}}}{\delta \psi_i(t)} \biggr\lvert_{\bm{\psi}(t)  = 0} = \lim_{N\rightarrow\infty}\frac{1}{N}\sum_i\langle x_i(t)\rangle_* \lvert_{\bm{\psi}(t)  = 0}, \label{start_1} \\
0 & = \lim_{N\rightarrow\infty} \frac{-1}{N} \sum_i \frac{\delta\overline{\mathcal{Z}}[\bm{\psi}(t)  = 0]}{\delta h_i(t)} = \lim_{N\rightarrow\infty}\frac{-i}{N}\sum_i\langle \hat{x}_i(t)\rangle_*, \\
C(t,t') & = \lim_{N\rightarrow\infty}\frac{-1}{N}\sum_i \frac{\delta^2\overline{\mathcal{Z}}}{\delta \psi_i(t)\delta \psi_i(t')}\biggr\lvert_{\bm{\psi}(t)  = 0} = \lim_{N\rightarrow\infty}\frac{1}{N}\sum_i\langle x_i(t)x_i(t')\rangle_* \lvert_{\bm{\psi}(t)  = 0}, \\
G(t,t') & = \lim_{N\rightarrow\infty}\frac{-i}{N}\sum_i \frac{\delta^2\overline{\mathcal{Z}}}{\delta \psi_i(t)\delta h_i(t')} \biggr\lvert_{\bm{\psi}(t)  = 0} = i\lim_{N\rightarrow\infty}\frac{1}{N}\sum_i\langle x_i(t)\hat{x}_i(t')\rangle_* \lvert_{\bm{\psi}(t)  = 0},\\
0 & = \lim_{N\rightarrow\infty}\frac{-1}{N}\sum_i \frac{\delta^2\overline{\mathcal{Z}}[\bm{\psi}(t)  = 0]}{\delta h_i(t)\delta h_i(t')} = \lim_{N\rightarrow\infty}\frac{1}{N}\sum_i\langle \hat{x}_i(t)\hat{x}_i(t')\rangle_* . \label{end_1}
\end{align}
Notice that since $\mathcal{Z}[\bm{0}, \bm{h}(t)] = 1$ for every choice of $\bm{h}(t)$, those expressions that only have variations with respect to $h$ are zero. We can now set the fields $\bm{\psi}(t)$ to zero since they no longer have a purpose, and compare (\ref{start_op}-\ref{end_op}) with (\ref{start_1}-\ref{end_1}) to write the following identities at the physical saddle-point
\begin{equation}
a(t) = m(t), \qquad k(t)=0, \qquad Q(t,t')=C(t,t'), \qquad K(t,t')=-iG(t,t'), \qquad L(t,t')=0. \label{sp1}
\end{equation}

Variations with respect to the physical order parameters are calculated in the same way. For example, single-time variations result in
\begin{align}
    0 =& \frac{\delta}{\delta a(t)}\left[\Psi + \Phi + \Omega \right] = \frac{\delta \Psi}{\delta a(t)} = \frac{\delta}{\delta a(t)} \left(i\int ds\, \hat{a}(s)a(s)\right) \nonumber \\
    =& i\int ds \, \hat{a}(s)\frac{\delta a(s)}{\delta a(t)}  = i\int ds \, \hat{a}(s)\delta(s-t) = i\hat{a}(t) \ .
\label{saddle}
\end{align}
We also have
\begin{align}
    0 & = \frac{\delta}{\delta Q(t, t')}\left[\Psi + \Phi + \Omega \right] = i\hat{Q}(t,t') - \frac{\alpha v^2}{2c}L(t,t') - \nonumber \\
    & \hspace{1cm} \frac{\alpha c}{2}\frac{\int\mathcal{D}[u,v] u(t)u(t')e^{-\frac{1}{2}\int ds\, ds'\, \left[
	u(s)u(s')Q(s,s')+v(s)v(s')L(s,s') - 2v(s)u(s')K(s',s)\right] + i\int ds\, u(s)v(s)}}{\int\mathcal{D}[u,v] e^{-\frac{1}{2}\int ds\, ds'\, \left[
	u(s)u(s')Q(s,s')+v(s)v(s')L(s,s') - 2v(s)u(s')K(s',s)\right] + i\int ds\, u(s)v(s)}} \nonumber \\
    & = i\hat{Q}(t,t')-\frac{\alpha c}{2}\frac{\int\mathcal{D}[u,v] u(t)u	(t')e^{\int ds\, ds'\, \left[
	-\frac{1}{2}u(s)C(s,s')u(s') + iu(s)\left[\delta(s-s')-G(s, s')\right]v(s')\right]}}{\int\mathcal{D}[u,v] e^{\int ds\, ds'\, \left[
	-\frac{1}{2}u(s)C(s,s')u(s') + iu(s)\left[\delta(s-s')-G(s, s')\right]v(s')\right]}} \nonumber \\
& =  i\hat{Q}(t,t') + 0,
\end{align}
where in the third step we have substituted with the saddle-point equations \eqref{sp1}, and we have used standard Gaussian integrals such as the ones in  Eqs. (128)-(130) in \cite{coolen2000statistical}.

We further have
\begin{align}
    0 & = \frac{\delta}{\delta K(t, t')}\left[\Psi + \Phi + \Omega \right] \nonumber \\
    & =i\hat{K}(t,t') - i\alpha c\delta(t-t') - \frac{\alpha v^2\Gamma}{2c} \left[K(t',t)+K(t,t')\right] \nonumber \\
    &\quad +\alpha c\frac{\int\mathcal{D}[u,v] u(t)v(t')e^{-\frac{1}{2}\int ds\, ds'\, \left[
	u(s)u(s')C(s,s') - 2iu(s)\big[\delta(s-s')-G(s,s')\big]v(s')\right]}}{\int\mathcal{D}[u,v] e^{-\frac{1}{2}\int ds\, ds'\, \left[u(s)u(s')C(s,s') - 2iu(s)\big[\delta(s-s')-G(s, s')\big]v(s')\right]}}\nonumber \\
& = i\hat{K}(t,t') - i\alpha c\delta(t-t') + i \frac{\alpha v^2\Gamma}{2c}[G(t',t)+G(t,t')] +i\alpha c (\mathbb{I}-G)^{-1}(t',t).
\end{align} 
Finally, from the variation with respect to $L$ we obtain
\begin{align}
0 & = \frac{\delta}{\delta L(t, t')}\left[\Psi + \Phi + \Omega \right] \nonumber \\
& = i\hat{L}(t,t') - \frac{\alpha v^2}{2c}Q(t,t') -\frac{\alpha c}{2}\frac{\int\mathcal{D}[u,v] v(t)v(t')e^{-\frac{1}{2}\int ds\, ds'\, \left[
		u(s)u(s')C(s,s') - 2iu(s)\big[\delta(s-s')-G(s,s')\big]v(s')\right]}}{\int\mathcal{D}[u,v] e^{-\frac{1}{2}\int ds\, ds'\, \left[
		u(s)u(s')C(s,s') - 2iu(s)\big[\delta(s-s')-G(s, s')\big]v(s')\right]}} \nonumber \\
& = i\hat{L}(t,t')-\frac{\alpha v^2}{2c} C(t,t') - \frac{\alpha c}{2}\left[(\mathbb{I}-G)^{-1}C(\mathbb{I}-G^\dagger)^{-1}\right](t,t').
\end{align}

\medskip

Summarising, we have
\begin{equation}
\hat{a}(t) = 0, \qquad \hat{k}(t) = 0, \qquad \hat{Q}(t, t') = 0, \label{sp2}
\end{equation}
and 
\begin{align}
    &\hat{K}(t, t') = -\frac{\alpha v^2\Gamma}{c} G (t', t) - \alpha c[G(\mathbb{I}-G)^{-1}](t',t) \ , \nonumber \\ 
    &\hat{L}(t,t') = -i\frac{\alpha v^2}{2c} C(t,t') - \frac{\alpha ci}{2 }\left[(\mathbb{I}-G)^{-1}C(\mathbb{I}-G^\dagger)^{-1}\right](t,t'). \label{sp3}
\end{align}

\subsection{Obtaining the effective dynamics}

We have seen from Eqs. (\ref{start_1}-\ref{end_1}), that the magnitudes we are interested in can be obtained as averages over the measure $\langle \cdot\rangle_*$. If we substitute with the saddle-point equations \eqref{sp1}, \eqref{sp2}, and \eqref{sp3} the measure simplifies to
\begin{align}
\int \mathcal{D}[x,\hat{x}]p[x(0)]&\exp\left(i\int dt\, \hat{x}\left[\frac{\dot{x}}{x}-1+ux(t)+\sigma\zeta+h\right]\right) \nonumber \\
&\exp\left(-i\int dt\, dt'\, x(t)\left[-\frac{\alpha v^2\Gamma}{c} G (t', t) - c\alpha [G(\mathbb{I}-G)^{-1}](t',t)\right]\hat{x}(t')\right) \nonumber \\
&\exp\left(\int dt\, dt'\, \hat{x}(t)\left[\frac{\alpha v^2}{2 c} C +\frac{c\alpha }{2}\left[(\mathbb{I}-G)^{-1}C(\mathbb{I}-G^\dagger)^{-1}\right](t,t')\right]\hat{x}(t)\right).
\end{align} 

This is recognised as the generating functional of the following `mean-field' process,
\begin{equation}
\frac{\dot{x}(t)}{x(t)} = 1 - ux(t) - \sigma\zeta(t) - h(t) - \alpha \int dt'\, \left[\Gamma(1-c) G  + cG(\mathbb{I}-G)^{-1}\right](t,t') x(t') + \eta(t), \label{complete_ed}
\end{equation}
where
\begin{equation}
\langle\eta(t)\eta(t')\rangle = \alpha \left[c(\mathbb{I}-G)^{-1}C(\mathbb{I}-G^\dagger)^{-1}+(1-c) C\right](t,t').
\end{equation}

Notice that the perturbation field $h(t)$ and the dynamical noise $\eta(t)$ appear in the same way in Eq. \eqref{complete_ed}. We can simplify the equations by removing $h(t)$ and writing the response function as
\begin{equation}
	G(t,t') = \frac{\delta \langle  x(t)\rangle_*}{\delta h(t')} =  - \frac{\delta \langle  x(t)\rangle_*}{\delta \eta(t')}.
\end{equation}

\section{Fixed-point analysis}
\subsection{Fixed-point ansatz}
In addition to focusing on fixed points we make three main assumptions: time translation invariance (TTI), finite integrated response (FIR), and weak long-term memory (WLTM). We will follow mainly section 4.5 of \cite{coolen2005mathematical} to derive the consequences of these assumptions.

At fixed points, the following objects become constants,
\begin{equation}
	x(t) = x, \qquad C(t,t') = \langle x(t)x(t') \rangle_* = \langle x^2\rangle_* \equiv q, \qquad \eta(t) = \sqrt{q}\Sigma z.
\end{equation}
As a consequence of TTI, we also have $G(t,t')=G(t-t')=G(\tau)$, and by causality this response function can only be non-zero for $t>t'$.

The second and third assumptions relate to the response function $G(t,t')$. We assume that the integrated response 
\begin{equation}
	\chi = \int_{0}^\infty d\tau\, G(\tau)
\end{equation}
will be finite (FIR). Notice that since the perturbation field $h(t)$ appears with a global minus sign in Eq. \eqref{complete_ed}, we expect that an increase in $h$ will decrease the value of $x$, implying $\chi<0$.

We then also have
\begin{equation}
\int dt'\, (\mathbb{I}+G)^{-1}(t-t') = \sum_n(-1)^n\int dt'\, G^n(t-t') = \sum_n(-1)^n\chi^n = \frac{1}{1+\chi}.
\end{equation}

The final condition, weak long-term memory, is related to the question of whether the system always reaches the same stationary state. WLTM is equivalent to the assumption that a fixed point is globally attractive, so any initial condition or perturbation is eventually ``forgotten" as the system evolves. For our purposes it is enough to note that simulations confirm that this is indeed the case in the fixed-point phase.

\medskip

With these assumptions we can make the following replacements,
\begin{equation}
	\int dt'\, f[G(t,t')] \rightarrow f(\chi), \qquad C(t,t')\rightarrow q,\qquad x(t) \rightarrow x, \qquad \eta(t)\rightarrow \sqrt{q}\Sigma z.
\end{equation}
From Eq.~(7) we then have at the fixed point,
\begin{equation}
	0 = x\left[1-ux-\sqrt{q}\Sigma z-\alpha x\left(c\frac{\chi}{1-\chi}+\Gamma(1-c)\chi\right)\right], \label{d1}
\end{equation}
and Eq. (8) for the correlation of the noise turns into the variance of the static random variable $z$
\begin{equation}
	\langle \eta^2\rangle = \alpha\left(c\frac{q}{(1-\chi)^2}+(1-c)q\right)\equiv q\Sigma^2 \label{d2}\ .
\end{equation} 

In Eq.~(9) averages over realizations of the dynamics $\langle\cdot\rangle_*$ turn into averages over the static random variable $z$, so we can express them using Gaussian integrals. We find
\begin{align}
    \chi &= \frac{1}{\sqrt{q}\Sigma}\left\langle\frac{\partial x}{\partial z}\right\rangle_*
    =\frac{1}{\sqrt{q}\Sigma}\int_{-\infty}^\infty \frac{dz}{\sqrt{2\pi}}e^{-\frac{z^2}{2}}\frac{\partial x}{\partial z}
    =\frac{1}{\sqrt{q}\Sigma}\int_{-\infty}^\Delta Dz\frac{-\sqrt{q}\Sigma}{u+\alpha \left(c\frac{\chi}{1-\chi}+\Gamma(1-c)\chi\right)}   \nonumber \\
    &= f_0(\Delta)\frac{-1}{u+\alpha \left(c\frac{\chi}{1-\chi}+\Gamma(1-c)\chi\right)}, \nonumber\\
    M &= \langle x\rangle_* = \frac{1}{u+\alpha \left(c\frac{\chi}{1-\chi}+\Gamma(1-c)\chi\right)} \int^\Delta_{-\infty} Dz (1-\sqrt{q}\Sigma z) \nonumber \\
    &= \frac{\sqrt{q}\Sigma}{u+\alpha \left(c\frac{\chi}{1-\chi}+\Gamma(1-c)\chi\right)} \int^\Delta_{-\infty} Dz (\Delta-z) = f_1(\Delta)\frac{\sqrt{\alpha q\left(\frac{c}{(1-\chi)^2}+(1-c)\right)}}{u+\alpha \left(c\frac{\chi}{1-\chi}+\Gamma(1-c)\chi\right)},\nonumber \\
    q &= \frac{1}{\left[u+\alpha \left(c\frac{\chi}{1-\chi}+\Gamma(1-c)\chi\right)\right]^2} \int^\Delta_{-\infty} Dz (1-\sqrt{q}\Sigma z)^2 \nonumber \\
    &= \frac{q\Sigma^2}{\left[u+\alpha \left(c\frac{\chi}{1-\chi}+\Gamma(1-c)\chi\right)\right]^2} \int^\Delta_{-\infty} Dz (\Delta-z)^2 = f_2(\Delta)\frac{{\alpha q\left(\frac{c}{(1-\chi)^2}+(1-c)\right)}}{\left[u+\alpha \left(c\frac{\chi}{1-\chi}+\Gamma(1-c)\chi\right)\right]^2}, \label{c5}
\end{align}
where
\begin{equation}
	f_n(\Delta) = \int Dz (\Delta-z)^n , \qquad  Dz = \frac{dz}{\sqrt{2\pi}}e^{-\frac{z^2}{2}}.
\end{equation}

The integrals $f_n$ can be written as
\begin{align}
f_0(\Delta) =& \frac{1}{2}\left[1+\text{Erf}\left(\frac{\Delta}{\sqrt{2}}\right)\right],\\
f_1(\Delta) =&\frac{1}{2}\left[e^{-\frac{\Delta^2}{2}}\sqrt{\frac{2}{\pi}}+\Delta\left(1+\text{Erf}\left[\frac{\Delta}{\sqrt{2}}\right]\right)\right], \\
f_2(\Delta) =& \frac{1}{2}(1+\Delta^2)\left[1+\text{Erf}\left(\frac{\Delta}{\sqrt{2}}\right)\right]+\frac{\Delta}{\sqrt{2\pi}}e^{-\frac{\Delta^2}{2}},
\end{align}
and we have the useful identity
\begin{equation}
f_2(\Delta)-f_0(\Delta) = \Delta f_1(\Delta). \label{f_identity}
\end{equation}

The order parameters in the fixed-point regime are determined by Eqs.~(19), which are equivalent to Eqs.~(\ref{c5}). We now describe how to solve these equations. The fully connected and dilute models are treated separately, because they result in qualitatively different solutions.

We next seek to obtain $\alpha, \chi, M$ and $q$ parametrically as a function of $\Delta$, for given values of $u,c$ and $\Gamma$.

\subsection{Fully connected system}
For $c=1$ Eqs.~(\ref{c5}) turn into
\begin{eqnarray}
f_0(\Delta) =& -\chi  \left(\dfrac{\alpha  \chi }{1-\chi }+u\right), \label{c1}\\
f_1(\Delta) =& \dfrac{M(1-\chi)}{\sqrt{\alpha q}}\left(\dfrac{\alpha  \chi }{1-\chi }+u\right), \label{c2}\\
f_2(\Delta) =& \dfrac{(1-\chi )^2}{\alpha}\left(\dfrac{\alpha  \chi }{1-\chi }+u\right)^2 \label{c3}.
\end{eqnarray}
We notice that the first and last equations only contain $\chi$ and $\alpha$. Dividing equation \eqref{c3} by the square of equation \eqref{c1} we find
\begin{equation}
\alpha = \frac{(1-\chi)^2f_0^2}{f_2\chi^2}.\label{c4}
\end{equation}
Substituting this back into Eq.~\eqref{c1} leads to
\begin{equation}
f_0 =-\chi  \left(\frac{(1-\chi)^2f_0^2}{f_2\chi^2}\frac{\chi}{1-\chi }+u\right) = (\chi-1)\frac{f_0^2}{f_2}-\chi u 
\implies  \chi = \frac{f_0+\frac{f_0^2}{f_2}}{\frac{f_0^2}{f_2}-u}=\frac{f_0(f_2+f_0)}{f_0^2-f_2u}.
\end{equation}

Substituting the expression for $\chi$ into Eq.~\eqref{c4} we then arrive at
\begin{equation}
\alpha = \left(\frac{1}{\chi}-1\right)^2\frac{f_0^2}{f_2}=\left(\frac{f_0^2-uf_2}{f_0(f_2+f_0)}-1\right)^2\frac{f_0^2}{f_2}=\left(\frac{f_2(u+f_0)}{f_0(f_2+f_0)}\right)^2\frac{f_0^2}{f_2}=\frac{f_2(u+f_0)^2}{(f_2+f_0)^2}.  
\end{equation}

To obtain $M$ and $q$  we use Eq.~\eqref{f_identity} and the fact that ${\Delta=(1-\chi)/\sqrt{\alpha q}}$ to rewrite Eq.~\eqref{c2} as
\begin{align}
    M &= \frac{f_1\sqrt{\alpha q}}{(1-\chi+\alpha\chi)} = \frac{f_1(1-\chi)}{(1-\chi+\alpha\chi)}\frac{1}{\Delta} \nonumber\\
    &= \frac{f_1}{(1+\alpha\frac{\chi}{1-\chi})}\frac{f_1}{f_2-f_0} = \frac{f_1^2\chi}{f_0(f_0-f_2)} = \frac{f_1^2(f_2+f_0)}{(f_0-f_2)(f_0^2-uf_2)},
\end{align} 
where we have used Eq.~\eqref{c1}. 

Dividing the square of Eq.~\eqref{c2} by Eq.~\eqref{c3} we further obtain
\begin{equation}
q = M^2\frac{f_2}{f_1^2} = \frac{f_1^2 f_2 (f_0 + f_2)^2}{(f_0 - f_2)^2 (f_0^2 - uf_2)^2}.  
\end{equation}

\subsection{General values of the connectivity ($c<1$)}
\label{appendix:invertingg}
\subsubsection{Relations for order parameters in the fixed-point regime}
Using the same reasoning we used in the fully connected case (dividing $f_2$ by $f_0^2$ using  Eqs.~(\ref{c5}) and simplifying), we can obtain a cubic expression for $\chi$,
\begin{align}
0=&f_0 (c (\Gamma -1) f_0-\Gamma  f_0-f_2) \nonumber \\
&+\chi  \left(f_0^2[c+2\Gamma-2 c \Gamma]+ 2 f_0 f_2[1-c] - f_2 u \right) \nonumber \\
&+\chi ^2 (c-1) \left(\Gamma  f_0^2+f_0 f_2-2 f_2 u\right) \nonumber  \\
&+ \chi ^3 uf_2  (c-1). \label{cubic}
\end{align}

The remaining order parameters, and $\alpha$, can be expressed as functions of $\chi$
\BE
 	M &=& \chi\frac{f_1^2}{f_0(f_0-f_2)}, \nonumber \\ 
	q &=& \chi^2 \left(\frac{f_1}{f_2-f_0}\right)^2\frac{f_2}{f_0^2}, \nonumber \\
		\alpha &=& \frac{f_0^2}{f_2} \frac{1}{\chi^2(1-c+\frac{c}{(1-\chi)^2})}.
\EE

\subsubsection{Selection of the physically meaningful solution for $\chi$}
For a given value of $\Delta$ (and fixed model parameters $c, \Gamma$ and $u$) we  obtain $\chi$ as one root of the above cubic equation (\ref{cubic}).

\begin{figure}[h]
	\centering
	\includegraphics[width=1\linewidth]{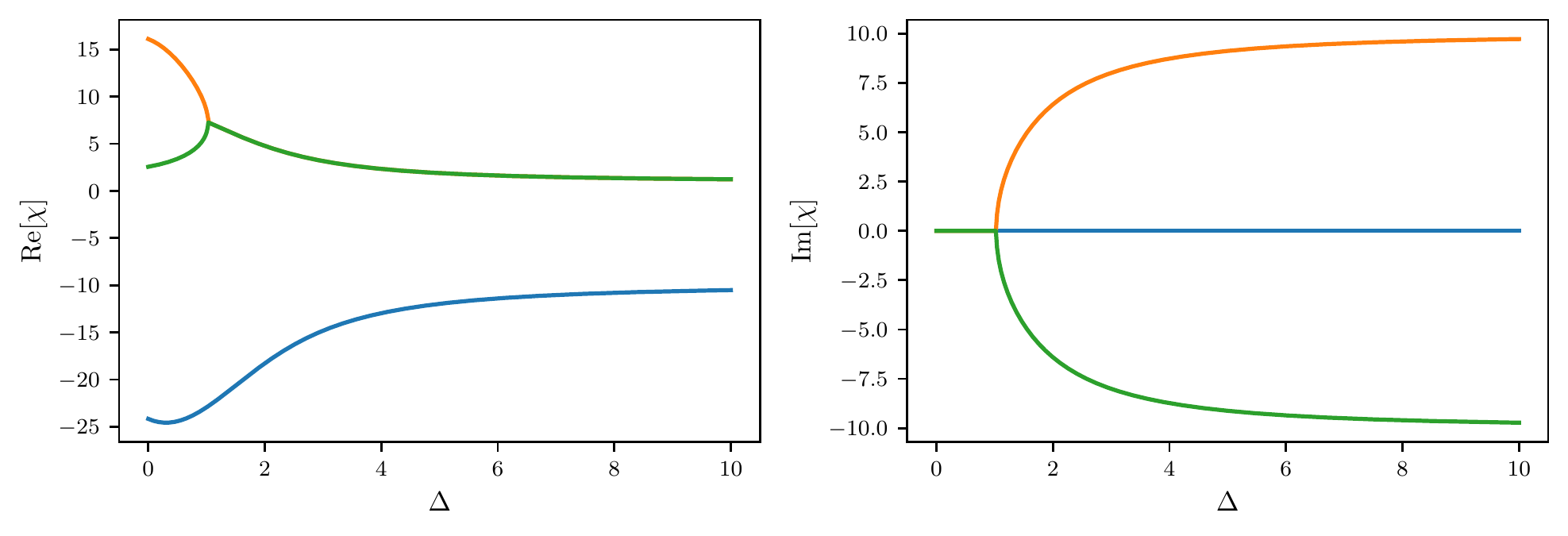}
	\caption{Real and imaginary parts of the roots of Eq. \eqref{cubic} for $u=0.1$, $c=0.99$, and $\Gamma=0.5$.}
	\label{fig:chig}
\end{figure}

Fig. \ref{fig:chig} shows how the solutions of Eq. \eqref{cubic} look when three real solutions exist. Notice that only one root is negative, as suggested by the fact that $\eta$ appears in the equations with a global negative sign, and that it is continuously connected with the case with a single real solution. 

The regions with three real solutions can be found by calculating the discriminant of the cubic polynomial in Eq.~\eqref{cubic}. When the discriminant is positive (and the coefficients are real) there are three real roots, if it is negative there is only one real root and a further pair of complex conjugate solutions.

\begin{figure}[h]
	\centering
	\includegraphics[height=0.3\linewidth]{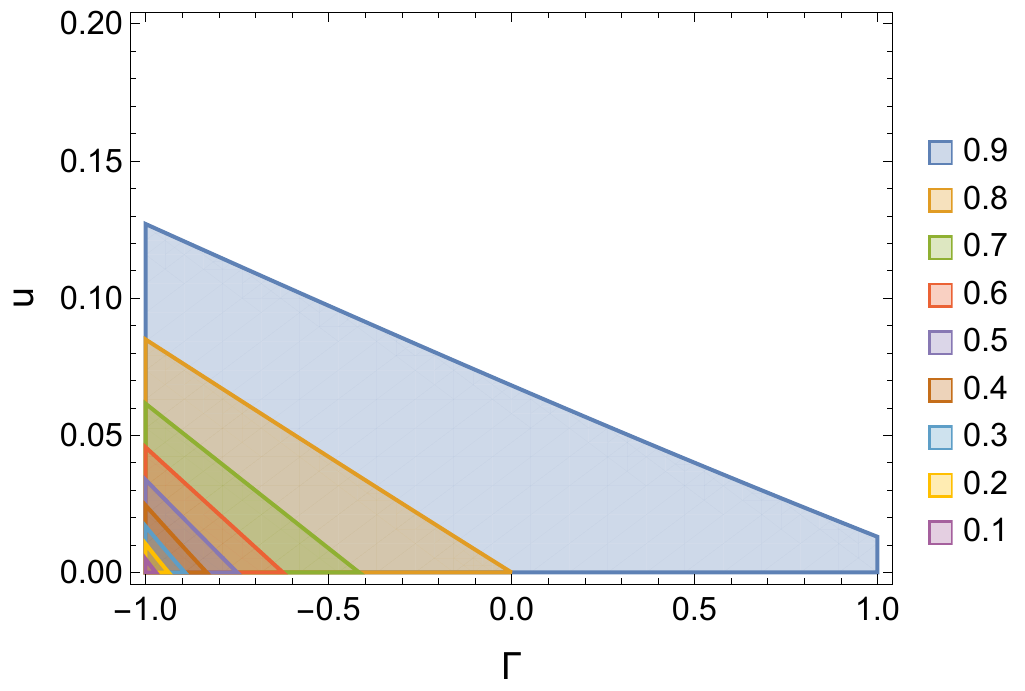}\hfil
	\includegraphics[height=0.3\linewidth]{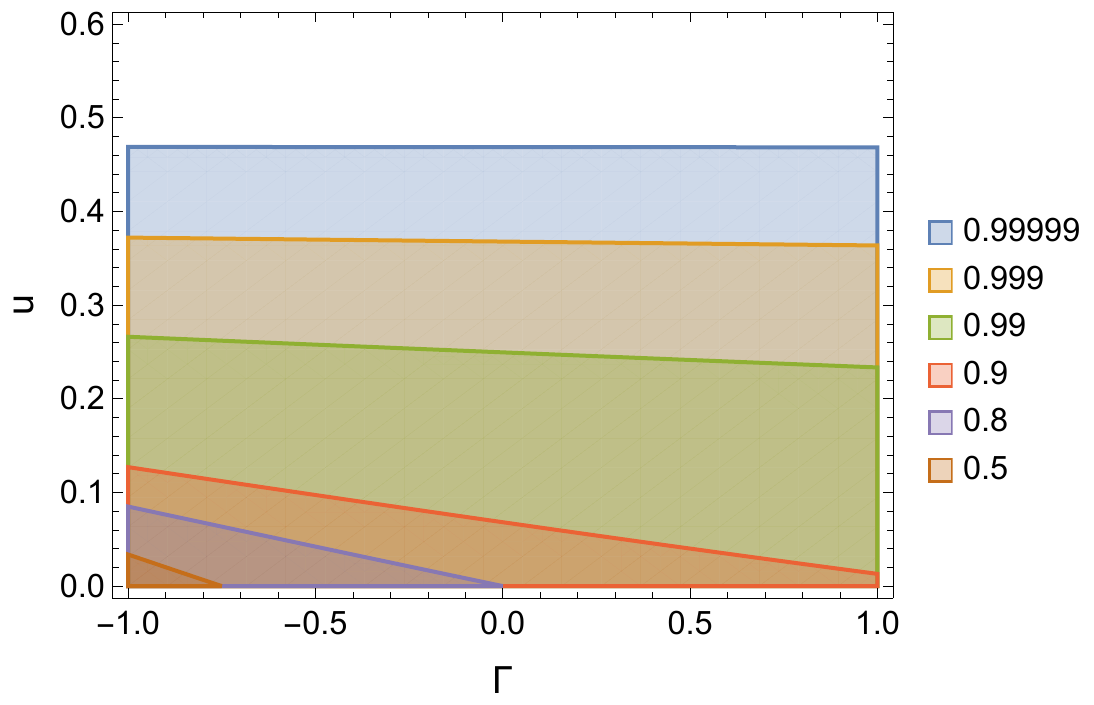}
	\caption{Parameter regions with positive values of the discriminant at $\Delta=0$. The different colours represent different values of $c$. Remember that, as discussed in Appendix \ref{Appendix:Bernoulli}, not all pairs of $(c, \Gamma)$ are possible.}
	\label{fig:regions}
\end{figure}

However, while the discriminant is easily calculated, it results in a very long equation, not suitable for direct analysis. We therefore investigate this numerically. Fig.~\ref{fig:regions} shows the regions where the discriminant is positive for $\Delta=0$ (this is where Fig.~\ref{fig:chig} suggests real solutions will first appear).

The right-hand panel of Fig.~\ref{fig:regions} indicates that, close to $c=1$, the condition to have three real solutions is independent of $\Gamma$. This can be further confirmed by Taylor expanding the discriminant to first order in $c$ about $c=1$. This results in $4 f_2 u (uf_2-f_0^2)^3 (c - 1)$, which is equal to zero whenever $uf_2-f_0^2=0$, that is, at the phase boundary for $c=1$ and $u<1/2$. For values of $c$ that are not close to one we find that three solutions only exist when $u$ is small (left-hand panel of Fig.~\ref{fig:regions}).

The takeaway of all this is the following: There exists a small region of parameter space, close to the line where $\chi$ diverges for $c=1$, where Eq.~(21) has three real solutions. Of these three solutions only one is always negative; this branch connects continuously with the regions that have a single real solution. We will only use this value of $\chi$.

\subsection{Solution for remaining order parameters}
Once $\chi$ is known we can substitute into
\begin{equation}
\alpha = \frac{f_0^2}{f_2}\frac{1}{\chi^2(1-c+\frac{c}{(1-\chi)^2})}  
\end{equation}
to obtain $\alpha$.

Once $\alpha$ and $\chi$ are known, we can proceed to find $M$ and $q$. From Eq. \eqref{f_identity} we obtain
\begin{equation}
q = \left(\frac{f_1}{f_2-f_0}\right)^2\frac{1}{\alpha\left(1-c+\frac{c}{(1-\chi)^2}\right)} = \chi^2 \left(\frac{f_1}{f_2-f_0}\right)^2\frac{f_2}{f_0^2},  
\end{equation}
and thus, using
\begin{equation}
q = M^2\frac{f_2}{f_1^2},  
\end{equation}
we have
\begin{equation}
M = \chi\frac{f_1^2}{f_0(f_0-f_2)}.
\end{equation}

\subsection{The limit $c\uparrow 1$}

We briefly discuss the limit $c\uparrow 1$. The observable order parameters $M, q$ and $\chi$ are continuous in this limit. However, the fraction of survivors can show a discontinuity.

Indeed, for $c<1$ the divergence always occurs when $f_0(\Delta)=f_2(\Delta)$, that is $\Delta=0$ and $\phi(\Delta=0)=f_0(\Delta=0)=1/2$. However, we have seen that for the case $c=1$, $u<1/2$,  the divergence occurs when $f_0^2-uf_2=0$ which implies $\Delta > 0$ and $\phi>1/2$.

\begin{figure}[h]
	\centering
	\includegraphics[width=0.6\linewidth]{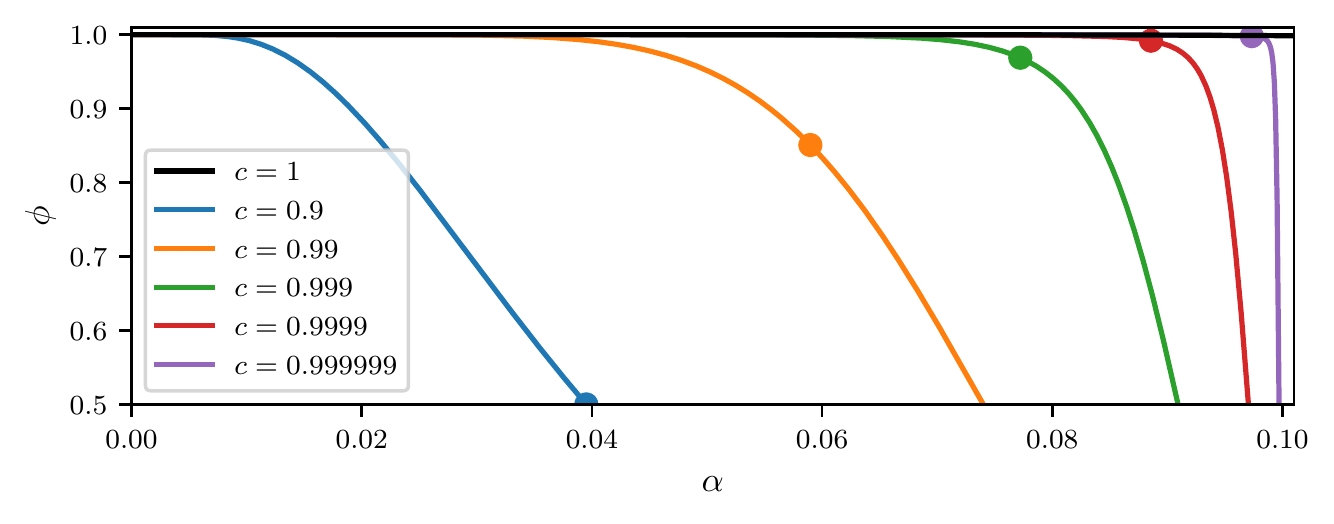}
	\caption{Convergence in $c$ of the fraction of survivors for $\Gamma=0.5$ and $u=0.1$ to the case $c=1$. The dots represent the points where three real solutions start to appear (the discriminant becomes zero).}
	\label{fig:nonuniform}
\end{figure}

As a result, we have $\phi(\alpha_c)=1/2$ for $u>1/2$, but if $u<1/2$ we have $\lim_{c\uparrow 1} \phi[\alpha_c(u, c, \Gamma)] \neq \phi[\alpha_c(u, c=1, \Gamma)]$. Fig. \ref{fig:nonuniform} provides some more insight on how this limit takes place. The curve for the fraction of survivors has pointwise, but not uniform, convergence to the case $c=1$.

As a final note we consider the curve $\phi(\alpha)$ in the intervals $[0, \alpha_s(u, c, \Gamma)]$ in which there is only a single real solution for $\chi$ for $c<1$ ($\alpha_s$ marks the onset of multiple solutions). We find continuous convergence to the curve $\phi(\alpha)$ for the fully connected model. This can be seen from the Taylor expansion of the discriminant of \eqref{cubic}, and numerically in figure Fig. \ref{fig:nonuniform}.

\bibliography{refs}